\documentclass[conference]{IEEEtran}
\usepackage{times}
\usepackage{balance}
\usepackage{url}
\usepackage{epsfig}
\usepackage{booktabs}
\usepackage{graphicx}
\usepackage{subfigure}
\usepackage{amsmath}
\usepackage{tabu}
\usepackage{amssymb} 
\usepackage{amsfonts}
\usepackage{color}
\usepackage{multirow}
\usepackage{algorithm}
\usepackage{algpseudocode}
\usepackage[utf8]{inputenc}
\usepackage[T1]{fontenc}
\usepackage{capt-of}
\usepackage{enumerate}
\usepackage{hyperref}
\usepackage[hyphenbreaks]{breakurl}
\makeatother

\newcommand{\para}[1]{{\vspace{0pt} \bf \noindent #1 \hspace{6pt}}}
\newcommand{\spara}[1]{{\vspace{0.5pt} \bf \noindent #1 \hspace{6pt}}}

\newcommand{\npara}[1]{{\vspace{-1pt} \em  #1 \hspace{6pt}}}

\newcommand{\readmore}[1]{{\color{black} #1}}
\newcommand{\shepherd}[1]{{\color{black} #1}}

\newcommand{\yzedit}[1]{{\color{black} #1}}

\newenvironment{packed_itemize}{
\begin{list}{\labelitemi}{\leftmargin=0.8em}
  \setlength{\itemsep}{3pt}
  \setlength{\parskip}{0pt}
  \setlength{\parsep}{0pt}
  \setlength{\headsep}{0pt}
  \setlength{\topskip}{0pt}
  \setlength{\topmargin}{0pt}
  \setlength{\topsep}{0pt}
  \setlength{\partopsep}{0pt}  
}{\end{list}}

\newfont{\mycrnotice}{ptmr8t at 7pt}
\newfont{\myconfname}{ptmri8t at 7pt}

\hypersetup{pdfstartview=FitH,pdfpagelayout=SinglePage}

\makeatletter
\def\@copyrightspace{\relax}
\def\@copyrightmode{\relax}

\begin{document}

\title{Et Tu Alexa? When Commodity WiFi Devices Turn into Adversarial Motion Sensors}

\author{
\IEEEauthorblockN{Yanzi Zhu$^\dag$, Zhujun Xiao$^*$, Yuxin Chen$^*$, Zhijing Li$^\dag$, Max
  Liu$^*$, Ben Y. Zhao$^*$, Haitao Zheng$^*$}
\IEEEauthorblockA{
$^\dag$University of California, Santa Barbara: \{yanzi, zhijing\}@cs.ucsb.edu\\
$^*$University of Chicago: \{zhujunxiao, yxchen, maxliu, ravenben, htzheng\}@cs.uchicago.edu
}
}

\IEEEoverridecommandlockouts
\makeatletter\def\@IEEEpubidpullup{6.5\baselineskip}\makeatother
\IEEEpubid{\parbox{\columnwidth}{
    Network and Distributed Systems Security (NDSS) Symposium 2020\\
    23-26 February 2020, San Diego, CA, USA\\
    ISBN 1-891562-61-4\\
    https://dx.doi.org/10.14722/ndss.2020.23053\\
    www.ndss-symposium.org
}
\hspace{\columnsep}\makebox[\columnwidth]{}}

\maketitle

\begin{abstract}

Our work demonstrates a new set of  silent 
    reconnaissance attacks, which leverages the presence of commodity WiFi
  devices to track users inside private homes and
  offices, without compromising any WiFi network, data
    packets, or devices.
  We show that just by sniffing existing WiFi signals, an 
  adversary can accurately detect and track movements of users
  inside a building. This is made possible by our new signal 
    model that links together human motion near WiFi transmitters and
    variance of multipath signal propagation seen by the attacker
    sniffer outside of the property.
  The resulting attacks 
  are cheap, highly effective, and yet difficult to detect.  We implement
  the 
  attack using a single commodity smartphone, deploy it in 11
  real-world offices and residential apartments, and show it is 
  highly effective. Finally, we evaluate potential defenses, and
  propose a practical and 
  effective defense based on AP signal obfuscation.

\end{abstract}

\section{Introduction}
With near-ubiquitous deployment of WiFi-enabled smart devices ({\em
  e.g.}, security cameras, voice assistants, and smart appliances),  our
homes and offices are filled with many WiFi
devices\footnote{The worldwide number of WiFi-enabled IoT devices is expected to
  reach 5 billion by 2025~\cite{iot-wifi-number}, and the
  number of WiFi connected devices will reach 22.2 billion by
  2021~\cite{wifi-number}.}. The ubiquity of these devices and their sheer
density means that they will fill the air around us with radio frequency (RF)
signals, wherever we go. 

Unfortunately, the RF signals emitted by these devices pose a real security and privacy risk to all of us. They are constantly
interacting with ({\em e.g.}, reflecting off) our bodies, carrying 
information about our location, 
movement and other physiological properties to anyone nearby with sufficient
knowledge and curiosity.  In this work, we explore a new set of
passive reconnaissance attacks  that
leverages the presence of  {\em ambient WiFi signals} to monitor 
users in their homes and offices, 
even when the WiFi network, data packets, and individual
devices are completely secured and operating as expected.  We show that by just
sniffing existing WiFi signals, an adversary outside of the
  target property can accurately detect
and track movements of any users down to their individual rooms, regardless
of whether they are carrying any networked devices.

We believe this is the first in a new class of silent 
    reconnaissance attacks that are notable
because of their passive nature and general applicability. This attack can be
highly effective, incurs low cost (only cheap commodity hardware), and yet
remains {\em undetectable}.  The 
attacker does not need to compromise/access the WiFi network or individual
devices, decode packets or transmit any signals.  All they need is to place a
single, off-the-shelf, minimally equipped WiFi receiver outside of the target
property.  This attacker receiver only needs a {\em single}
antenna, and simply measures the signal strength of existing WiFi signals,
without decoding any packets.

Unaddressed, these reconnaissance attacks put our security and privacy at
significant risk.  \shepherd{The ability for an attacker to continuously and automatically
  scan, detect and locate humans behind walls at nearly no cost and zero risk ({\em
    e.g.\/} attacker waits for notifications remotely) will enable
  attackers to launch strong physical attacks and commit serious crimes. Such
  threat broadly applies to our homes, businesses, government
  facilities and many others.  Examples include burglary to homes and
    offices, kidnapping and assault of targets in their homes, ``casing'' a
    bank prior to robbery,  and even planning attacks against government
    agencies. 
}

\para{Why WiFi sensing?}  We note that there are some simple approaches to
inferring user presence that do not require the use of sophisticated
RF sensing. 
\shepherd{For example, attackers could infer user presence by
observing lighting or acoustic conditions inside an area, or use thermal imaging. These
attacks are well understood and easily disrupted by time-controlled lighting or sound 
systems~\cite{smartlighting}, or insulated walls designed to
prevent heat leakage and naturally block thermal
imaging~\cite{thermalimaging}. } Finally, attackers can infer user presence
from increased WiFi network traffic. Yet this is highly unreliable, as growth of
IoT devices increases traffic levels in the absence of users. It is also easily
thwarted using cover traffic~\cite{covertraffic}.

Instead, we describe a new class of physical reconnaissance attacks enabled by
  inherent properties of WiFi signal propagation: 
1) user movement near a WiFi transmitter changes its signal propagation in a
way that can be observed by nearby receivers, and 2) walls and buildings
today are not built to insulate against WiFi signals, thus signals sent by
devices inside a property can often be overheard by outside
receivers.  \readmore{Leveraging these, we design the attack such
  that}, whenever a WiFi device transmits signals, it unknowingly turns into a
tracking device for our attack.  In this context, our attack could
be viewed as an adversarial analogy to WiFi-based device-free human sensing
({\em e.g.}, see-through-wall systems that actively transmit
  customized RF signals towards the target~\cite{adib2013}).  Yet
  our attack differs significantly (\S\ref{sec:bksensing}), because we use a novel model on
  multipath signal
  dynamics to remove dependence on active
  transmissions (only passive sensing), customized hardware (only a
  commodity, single antenna receiver), \readmore{and knowing precise locations of WiFi devices 
  inside the property.}

\para{Motion detection via multipath signal dynamics.} The core of our attack is a new model on signal dynamics that links
  together human motion near WiFi transmitters and variance of
  multipath signal propagation seen by a sniffer outside of
  the property. Specifically, when a human target moves ({\em e.g.}, sitting
down, walking, opening/closing doors) near a WiFi device
$x$,  the motion changes the multipath signal propagation from $x$ to
the attacker sniffer $S$.  Our new signal
model allows $S$ to accurately capture such signal dynamics and use
them to pinpoint the target to her specific room. The more WiFi devices inside the
property, the more accurate the tracking becomes.  

Our proposed attack does not assume any prior knowledge of the WiFi network
and devices inside the target property, \readmore{including their 
locations. Our attack can discover 
devices and estimate their coarse locations using their WiFi signals}, and
the attack continues to function even if these devices are
relocated. 

We build a complete prototype of the 
attacker system on commodity smartphones, and experimentally show that the attack
(using a single smartphone) is not only highly accurate (detecting and
localizing users to an individual room), but also highly general (effective
across a wide range of 11 different physical settings, including both office buildings 
and residential apartments).

\para{Defense via AP-based signal obfuscation.}  We explore
robust defenses against our proposed attack and other passive 
sensing attacks. 
We consider four immediate defenses:
reducing leakage by geo-fencing and rate limiting, signal obfuscation
by MAC randomization, and power randomization at WiFi devices, and find that
they are either impractical or ineffective.  We then propose a practical alternative
using {\em AP-based signal obfuscation},
where the WiFi Access Point actively injects customized cover signal for its
associated devices.  This defense effectively creates noise to the signal
measurements, such that the attacker is unable to identify change due to
human motion.  Our defense is easy to
implement, incurs no changes to devices other than the AP, but reduces
the human detection rate to 47\% while increasing the
false positive rate to 50\%. Such ambiguity renders the attack
useless in practice.

In the rest of the paper, we describe our efforts to understand the feasibility,
challenges, and defenses surrounding the proposed attack.  In
short, our key contributions include: 
\begin{packed_itemize} 
\item  We identify a low-cost, undetectable human sensing attack using
  just a single sniffer with a single antenna, and design a new
  multipath signal variance model for motion detection. 
  
\item We prototype the attacker system on a commodity smartphone
  and validate the attack in real-world settings. 

\item We propose and evaluate a practical and effective
  defense using AP-based signal obfuscation. 
\end{packed_itemize}

\para{Limitations.} 
\shepherd{Currently, our attack detects 
human presence in each room over time}
by detecting and localizing targets to individual rooms.  It is unable
to identify fine-grained features such as speed, activity
type and user identity, or separate humans from large animals. 
Despite such limitations, our work identifies a realistic, low-cost,
and undetectable 
reconnaissance attack using passive WiFi sensing.  We hope our work brings more 
attention to this important and understudied topic.

\begin{figure}[t]
\centering
  \includegraphics[width=0.42\textwidth]{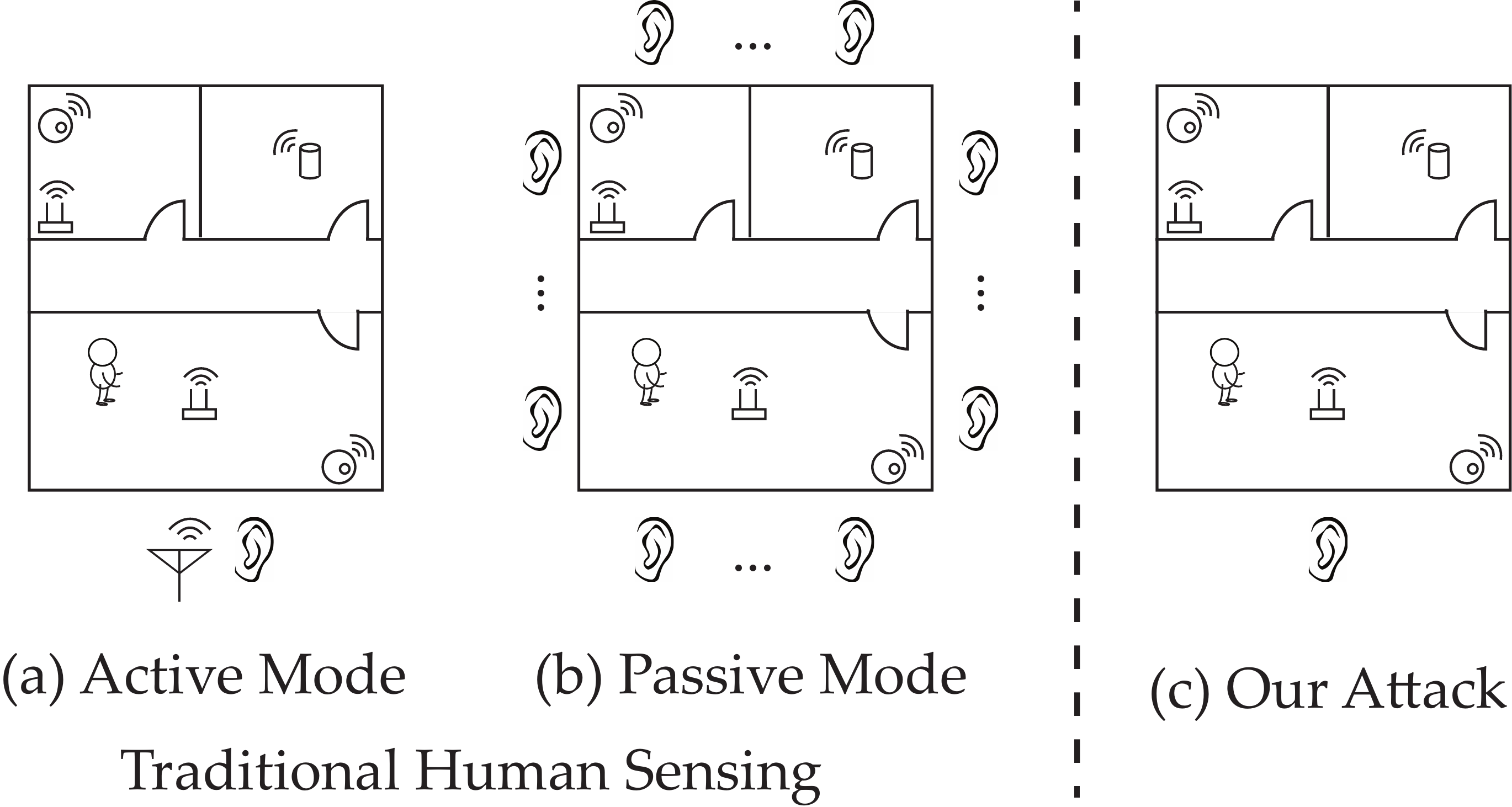}
\caption{Traditional human sensing designs either (a) relies on active
  transmissions by (customized) attacker devices, or (b) deploys one
  or more advanced sniffers (laptops/USRPs) with multiple antennas;  (c) Our attack uses
  a single smartphone (with a
  single antenna) as the passive sniffer, and turns commodity WiFi devices
  inside the property as motion sensors. 
}
\label{fig:mov}
\end{figure}

\section{Background: Device-free Human Sensing} 
\label{sec:bksensing}

Some details of adversarial sensing attacks are reminiscent of the problem of
``device-free human sensing.''  A natural question is: {\em can we simply
  reuse existing work on device-free human sensing systems to launch
  adversarial sensing attacks}?  To answer this question, and to better
understand how these attacks in the context of prior work, we review in detail
existing works in device-free human sensing.

The task of ``device-free human sensing'' makes no assumptions
on whether targets are carrying networked devices. Sensing is generally
achieved by capturing and analyzing RF signals reflected off or blocked by
human bodies. To be clear, this is quite different from the task of ``device
localization,'' in which the target is a networked device that
communicates and synchronizes 
with the sensing system, {\em i.e.\/} sending and/or receiving signals ({\em
  e.g.},~\cite{arraytrack13nsdi,spotfi15sigcomm,pinpoint06mobisys,
  radar00victorbahl,localization13survey}).

Existing works on device-free human sensing can be categorized into
two broad groups:  {\em active mode} and {\em passive mode}.

\para{Active sensing.}  Most of the existing works fall into this
group, where the sensing device continuously transmits RF
signals towards the target area (Figure~\ref{fig:mov}a).   As some signals get 
reflected off the target body,  they are captured by the sensing
device to infer the target status ({\em
  e.g.},~\cite{adib2013,adib15nsdi,lifs16mobicom, tsinghua14mobicom}).  To
facilitate sensing/inference, the RF signals are often custom-designed
to capture information of the target, {\em e.g.}, frequency-modulated continuous
wave (FMCW)  signal~\cite{adib2013,adib15nsdi} that largely differs from WiFi
transmissions.  We note that some prior works on active sensing  ({\em
    e.g.},~\cite{widet18mswim,widar2_18mobisys,youssef2007challenges})
 are branded as ``passive sensing'' to refer to device-free human
  sensing, although their 
  sensing device is actively transmitting signals. 

When considering our adversarial scenario in the context of active sensing,
the key property is ``detectability.''  Since 
the attacker device must {\em continuously} transmit RF
signals,  it is easy to detect, localize and remove these devices.

\para{Passive sensing.}  In a passive mode, sensing devices only listen to
existing transmission signals, but do not transmit signals themselves. They
have no control of the RF signal used for sensing. The state-of-the-art
design~\cite{lifs16mobicom} deploys multiple sniffers to listen to WiFi
signals sent by multiple transmitters in the target
area, and uses these signals to detect and
estimate user location. Specifically, when a user blocks the direct line of
sight (LoS) path from a transmitter to a sniffer, the original signal will diffuse
around the user.  By building a precise propagation model on signal diffusion
on the LoS path, \cite{lifs16mobicom} is able to derive the user location.
However, doing so requires precise location of the transmitters (cm-level).  Such
requirement is impractical under our adversarial scenario.

Similarly, an earlier work~\cite{banerjee14wisec} detects 
\shepherd{the presence of a user}
when she disturbs the direct path between a WiFi access point (AP) and a 
sniffer.  Again, the attacker must 
obtain AP locations a priori, and must deploy multiple
sniffers around the target area to detect \shepherd{user presence} 
(see Figure~\ref{fig:mov}b).

\para{Key observation.} While some existing human sensing systems can be
turned into attacks, they impose a hefty cost and risk for the attacker, significantly 
limiting the applicability of the attack. This motivates us to consider a new,
passive human sensing attack that can be launched by a minimally equipped
attacker and remains undetectable.   Along these lines, our proposed
attack only requires a single
commodity WiFi receiver (with a single antenna) outside of the target
area (Figure~\ref{fig:mov}c).  As we will explain in
\S\ref{sec:math_tripwire}, this is made possible by building a new
model to detect motion using dynamics of multipath signal propagation from 
each anchor to the sniffer, rather than those of the direct path as in~\cite{lifs16mobicom, banerjee14wisec}.

\section{Attack Scenario and Adversarial Model}

We start by describing the
attack scenario, the adversarial model, and the type of signals captured by the attacker sniffer.

\para{Attack scenario: one sniffer and many anchors.}  As shown in
Figure~\ref{fig:mov}c, our attack leverages the
ubiquity of commodity WiFi devices,  ranging from routers, desktops, printers, to IoT
devices like voice assistants, security cameras, and smart
appliances. These devices are often
spread over each room of our homes and offices~\cite{iotlayout2,iotlayout1}, and 
generally flood the surroundings with periodic WiFi
packets. We refer to these WiFi devices as {\em anchors} in our attack. 

Our attack also 
leverages the fact that WiFi signals are designed for significant coverage
and can penetrate most walls.  Thus an attacker can place a sniffer outside
the target property to passively listen to existing signals sent by
anchors, without compromising them or the network.  Because WiFi protocols do
not encrypt source and destination MAC addresses, the 
sniffer can isolate packets sent by each anchor, even under MAC
randomization~\cite{privacy14iccst, siby2017iotscanner,
  macrandomization}.

Our attack is effective if the sniffer can capture signals from at
least one anchor per room of interest.  The actual number of
sniffers required depends on the size and shape of the target property
and wall materials. Across all of our experiments with 11 
office buildings, residential apartments and single family houses
(described later in \S\ref{sec:locate_static}),  a single sniffer is
sufficient to cover our target scene. 

Our attack does not work when WiFi
signals do not leak to outside of the property, 
{\em e.g.\/} when the property has thick, concrete exterior walls.  The attacker can 
detect this (and walk away) when either the sniffer
sees too little WiFi signals, or the detected anchors are outside of the
target area (\S\ref{sec:locate_static}).

\para{Adversarial model.} We make the following assumptions about the adversary.  

\begin{packed_itemize}
\item The adversary makes no assumptions about the number, location, or
movement speed of human targets being tracked. 

  \item The adversary does not have physical or remote access to WiFi
  devices in the target property, or the property itself. 

\item Similar to the evil maid attack~\cite{evilmaid}, the attacker
  can physically move {\em outside} the target
  property, either outside exterior walls or along public corridors,
  without attracting suspicion.  This temporary access is necessary only for
  initial bootstrapping the attack, and not required for the prolonged sensing
  phase. 

\item To avoid detection, the attacker only performs passive WiFi sniffing,
  and avoids using any bulky or specialized hardware, {\em e.g.} directional antennas,
  antenna arrays, or USRP radios~\cite{usrp}. Instead, they use commodity
  mobile or IoT devices, {\em e.g.} smartphones or smart street lights. The
  sniffer device only needs a single (built-in) antenna.

  Note that while some smartphones (including the ones used in our attack
  implementation) have multiple antennas, their firmware only exposes 
  aggregate signal received across multiple antennas, effectively giving the
  same amount of information as devices with a single antenna.

\item \shepherd{The adversary partitions the target property into ``regions''
    or virtual rooms around the anchors to detect user presence. When the
    adversary has access to rough floor plans of the target
    property\footnote{\shepherd{Rough floor plan can often 
        be derived from publicly available data, such as real estate
        websites and apps, and public building
        documents.}}, the attacker detects user presence down to their
    individual rooms.}

\end{packed_itemize}

We intentionally choose a resource-limited attacker to demonstrate 
the generality of this attack.  Lower resource
requirements imply that the attack can be successful in a wider range
of practical scenarios.

\para{Signals captured by the sniffer.} For each anchor $x$, the
sniffer $S$ can extract two
metrics from $x$'s raw WiFi signals (even if the packets are
encrypted). The first is {\em amplitude of channel state
  information (aCSI)} that measures the received signal strength (RSS) on
each of the many frequency subcarriers used in a transmission. Since human
movements change the multipath signal propagation from $x$ to $S$,
$x$'s aCSI values seen by $S$ will fluctuate over time. The second one
is RSS, or the aggregated aCSIs over all the subcarriers. 
  This aggregation makes RSS relatively insensitive to human movements. 

A passive sniffer with a single antenna is
{\em unable} to extract advanced signal features including phase of
CSI (fCSI), Angle of Arrival
(AoA) and Time of Flight
(ToF)~\cite{wideo15nsdi,widar2_18mobisys}. Tracking fCSI and ToF requires the sniffer to actively synchronize with the
transmitter~\cite{vasisht2016decimeter},  and estimating AoA 
requires the sniffer to have an antenna array~\cite{arraytrack13nsdi,mostofi18ipsn}.  As mentioned earlier, while some 
smartphones are equipped with multiple antennas, their firmware only
reports a single effective CSI but not per-antenna CSI values.
Furthermore,   AoA estimation requires antennas to be
separated by half a wavelength (12.5cm for WiFi). Thus a
reasonable array of 4 antennas will be at least 19cm in width. 
These physical limitations rule out the feasibility of using phase, ToF and
AoA in our sensing design.

\begin{figure*}[t]
 \centering
 \subfigure[$\overline{\sigma_{aCSI}}$ captures user movement]{
   \includegraphics[width=0.32\textwidth]{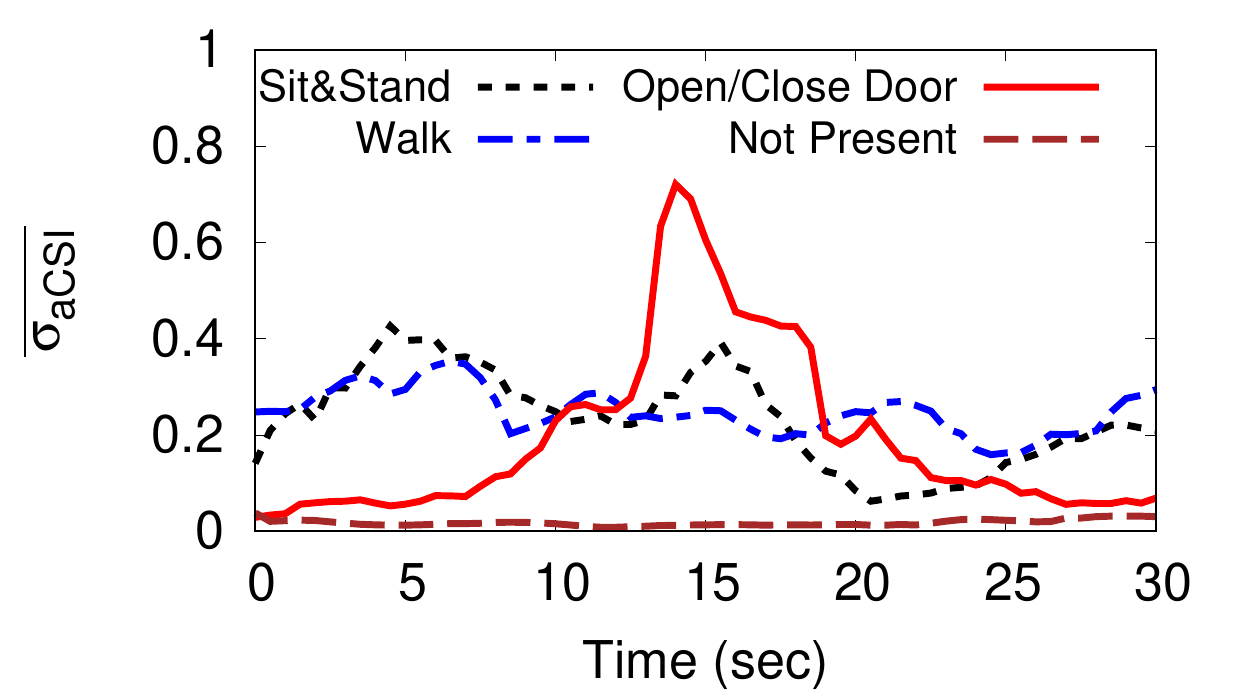}
} \hfill
\subfigure[$\overline{\sigma_{aCSI}}$ when the user is near, far from
the anchor, or completely absent]{
   \includegraphics[width=0.32\textwidth]{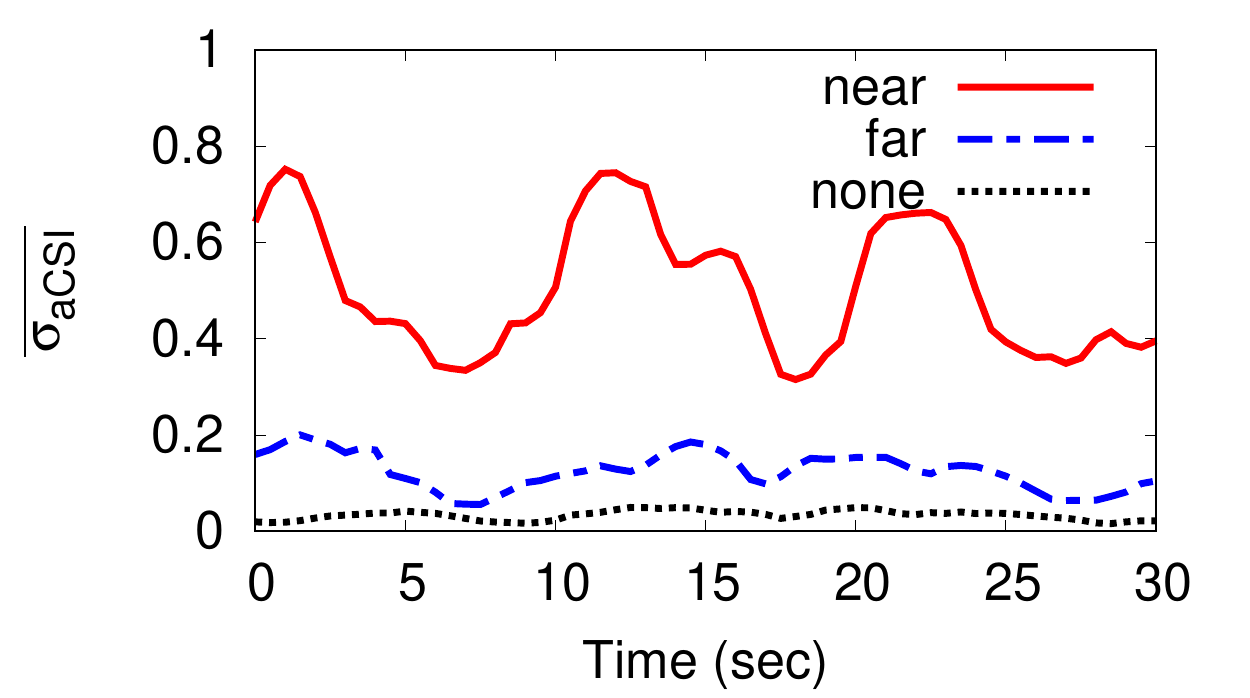}
   } \hfill 
\subfigure[Illustration of $\overline{\sigma_{aCSI}}$'s near-far phenomenon]{
   \includegraphics[width=0.28\textwidth]{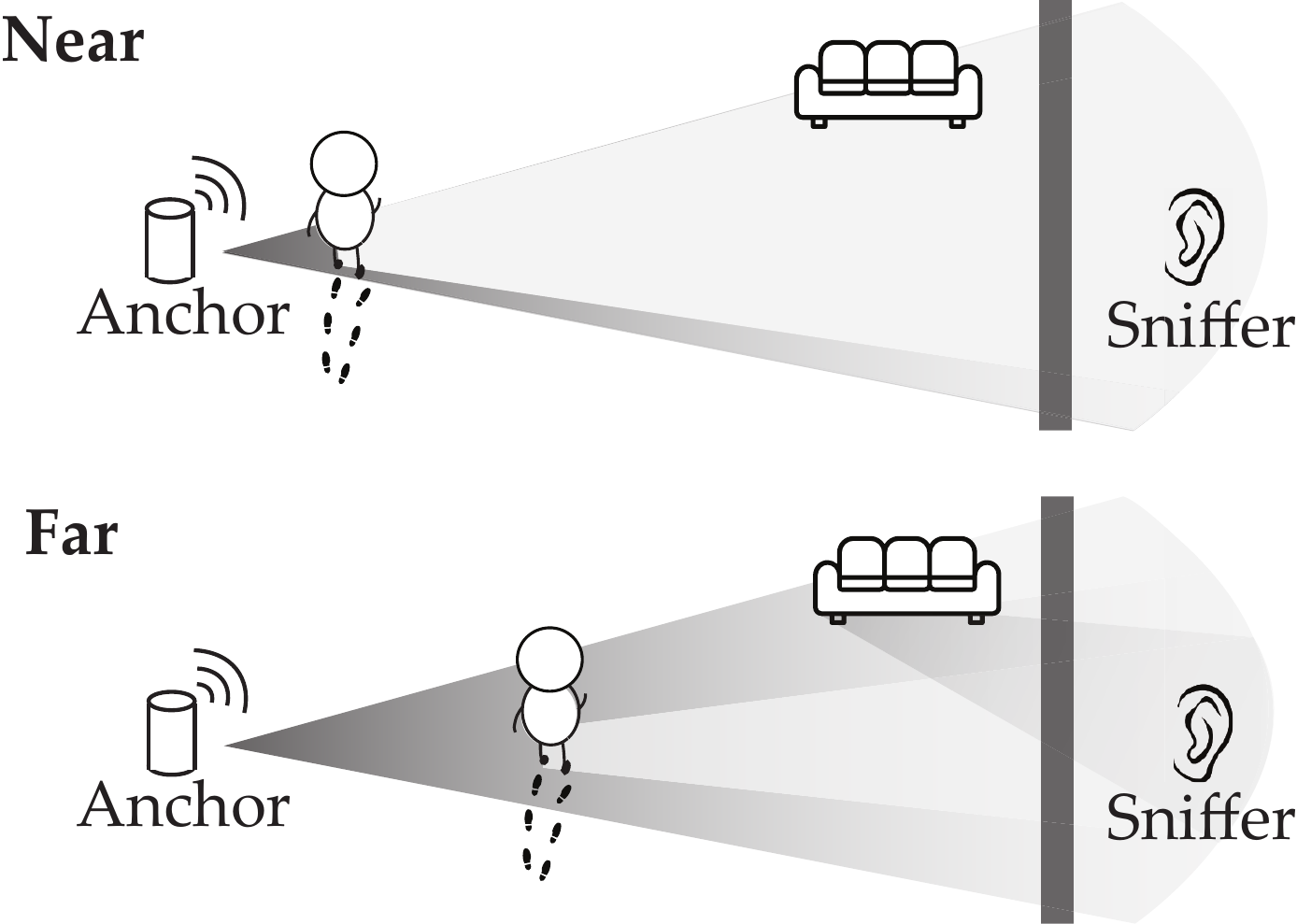}
  }
 \caption{Observations on how human movements affect an anchor's 
   $\overline{\sigma_{aCSI}}$ seen by the
   sniffer. 
 (a) $\overline{\sigma_{aCSI}}$ w/ and w/o user presence;
 (b)-(c)  When a user moves near an
  anchor $x$, some signal paths from $x$ to
  the sniffer are more frequently affected, so
  $\overline{\sigma_{aCSI}}(x)$ rises.
  As she moves away from $x$ and has
  less impact on the signal propagation, 
  $\overline{\sigma_{aCSI}}$ reduces.
}
 \label{fig:csi}
\end{figure*}

\vspace{-0.02in}
\section{Turning WiFi Devices into Motion Sensors}
\label{sec:math_tripwire}

Our attack is driven by a new aCSI variance model that links 
human motion near any anchor to temporal dynamics of the anchor's multipath signal 
propagation seen by the attacker sniffer. Whenever an anchor
transmits WiFi signals, it unknowingly turns into a motion sensor for
our attack.   These ``motion signals'' are immediately seen by the
attacker sniffer, who then pinpoints the targets down to their exact
room(s).

Unlike prior work on passive RF sensing~\cite{lifs16mobicom,
  banerjee14wisec}, our new model focuses on capturing temporal dynamics of
multipath signal propagation\footnote{WiFi signals sent by an anchor,
  when arriving at the sniffer, will go through rich multipath propagation, {\em
    e.g.}, reflections by furniture, walls and human.}  from each anchor to
the sniffer, rather than only the direct path.  This lets the attacker detect
any motion {\em around} each anchor that disturbs the multipath signal
propagation, and also eliminates the need to obtain precise
  anchor locations and deploy multiple sniffers~\cite{lifs16mobicom, banerjee14wisec}.

In the following, we describe the basic observations that motivate us
to pursue the attack, the key challenges it faces, 
and the key design concepts that make the
attack feasible.

\subsection{Correlation between Signal Dynamics and User Movement}
\para{(i) User movement $\rightarrow$ aCSI variance.} 
In an office/home, human users are never completely
stationary. Whether it is playing games, walking, opening doors,
sitting down, their natural movements will disturb the
multipath signal
propagation of nearby WiFi transmitters ({\em i.e.\/} anchors), creating
immediate, temporal 
variations in their aCSI values seen by the sniffer. 

We propose to capture such temporal variation by a new {\em aCSI
  variance} 
metric: 
\begin{equation} \label{eq:acsi} 
\overline{\sigma_{aCSI}} = \frac{1}{|I_q|} \sum_{i\in {I_q}}
\sigma_{aCSI}(f_i) 
\end{equation}
where  $\sigma_{aCSI}(f_i)$ represents the aCSI {\em standard deviation} for
subcarrier $i$ (at frequency $f_i$) calculated by the sniffer over a short 
time window ({\em  e.g.}, 5s).  We also take efforts to reduce the impact of noise
and artifacts in aCSI reports by the firmware,  first denoising 
aCSI per subcarrier using the wavelet method~\cite{passivecsi18NaNA},  then removing outliers by only including 
subcarriers whose  $\sigma_{aCSI}(\cdot)$ sequences are highly
correlated.  The set of subcarriers used in the above 
calculation ($I_q$) are the top 50\% of most correlated pairs. 

Figure~\ref{fig:csi}a plots several 30-second samples of
an anchor's $\overline{\sigma_{aCSI}}$ seen by the sniffer, 
for scenarios of no human presence,
a nearby user sitting down and standing up, opening/closing the door, and walking.  Compared
to no human presence, user movements lead to much higher
$\overline{\sigma_{aCSI}}$. 

We also find that user motion differs from equipment 
motion commonly seen in homes and offices,  {\em e.g.\/} an oscillating fan and a
robot vacuum. The latter either is too weak to produce any notable impact
on 
$\overline{\sigma_{aCSI}}$ or generates periodic signal patterns
different from those of human motion (\S\ref{sec:eval}).


\begin{figure*}[t]
 \centering
\includegraphics[width=0.98\textwidth]{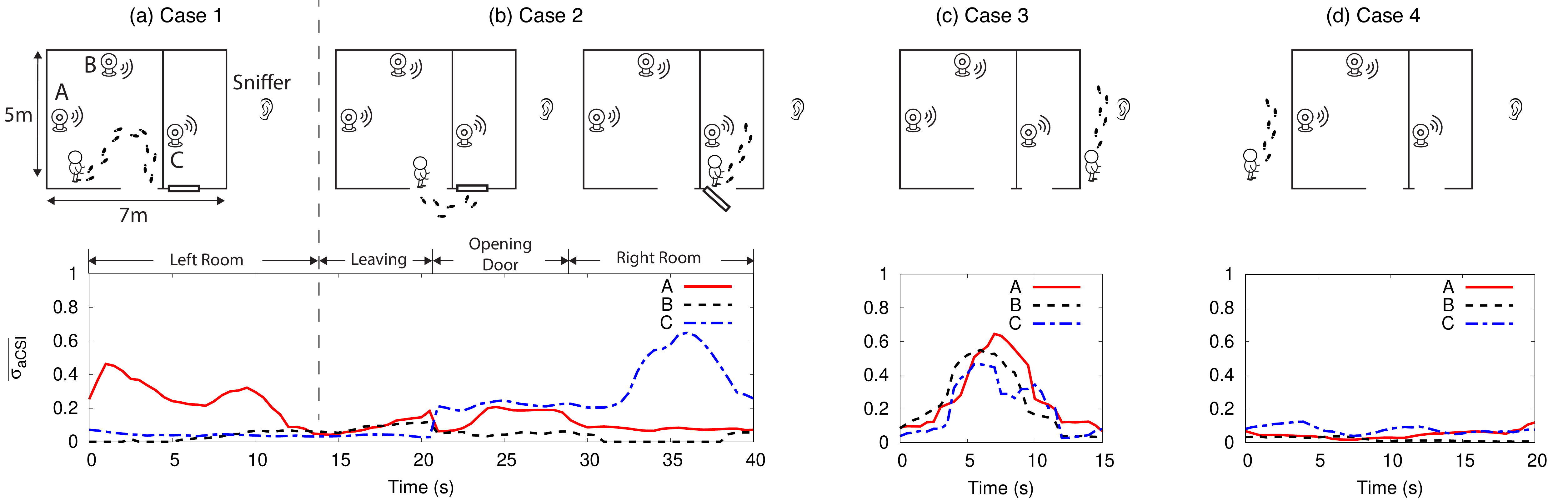}
 \caption{Four (simple) cases on user presence and the corresponding
   $\{\overline{\sigma_{aCSI}}\}$ traces from anchors A, B, and C.} 
 \label{fig:csi_single}
\end{figure*}

\begin{figure*}[t]
  \centering
  \includegraphics[width=1\textwidth]{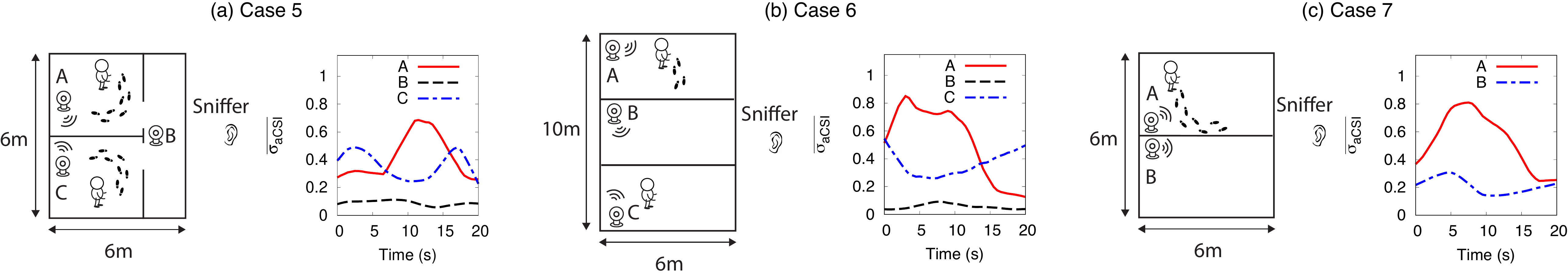}
  \caption{Three (complex) cases on user presence and the corresponding  $\{\overline{\sigma_{aCSI}}\}$ traces.}
  \label{fig:csi_multi}
\end{figure*}

\para{(ii) $\overline{\sigma_{aCSI}}$ depends on user-anchor
 distance.}   Another key observation is that when a target is
far away from an anchor $x$, its movements will
produce less disturbance to the signal propagation 
from $x$ to the
sniffer. This is demonstrated in
Figure~\ref{fig:csi}b, which compares an anchor
$x$'s $\overline{\sigma_{aCSI}}$ when a human user (walking) is close to $x$,
far from $x$ (in a different room), or completely absent.

We think this is due to the fact that  a target is ``bigger'' when it is closer
(Figure~\ref{fig:csi}c). As a target moves in the space between an anchor
$x$ and the sniffer, \yzedit{it} blocks and diffracts some signal paths from $x$ to
the sniffer. When close to $x$, \yzedit{it} affects more paths than when \yzedit{it}
is far away from $x$. Thus the received signals seen by the sniffer
will display a larger temporal variation when the user is closer to $x$. This
phenomenon can be modeled using an abstract, ray-tracing model on
$\overline{\sigma_{aCSI}}$ (omitted due to space
limits). Given a fixed time period, user
movements near $x$ create more path dynamics than those far from $x$,
leading to a larger standard deviation in the received signal strength (per
subcarrier).  

We validate this observation by measuring $\overline{\sigma_{aCSI}}$
of multiple anchors (Table~\ref{tbl:devices_summary})  in 11 test scenes
(Table~\ref{tbl:attack_environments}). As a target moves in
the space between an anchor and the sniffer, we see a general
tendency of 
$\overline{\sigma_{aCSI}}$ decreasing with the anchor-to-target
distance.  We experiment with different wall materials 
({\em e.g.}, glass, wood), distance of anchor and sniffer (8m--15m),
and sniffer placement ({\em e.g.}, on the floor, in the bush,
underneath a plastic trash can), and observe the same trend.

\para{(iii) $\overline{\sigma_{aCSI}}$ is a more robust motion
  indicator than $aCSI$.} 
Prior work~\cite{lifs16mobicom} localizes targets by 
modeling $aCSI$ of the direct path. This requires an accurate propagation model and the
precise physical location of each anchor. Instead, our 
$\overline{\sigma_{aCSI}}$ based method targets multipath
dynamics caused by user motion,  thus only
requires knowing the room each anchor resides, rather than its precise
location inside the room.

\subsection{Challenge: Sensing Ambiguity}
The above discussion suggests that with a sufficient number of anchors in a room,
the sniffer should be able to detect human motion in the room from its 
anchors' $\overline{\sigma_{aCSI}}$.  For example, if any
anchor's $\overline{\sigma_{aCSI}}$ is sufficiently large, {\em
  i.e.\/} motion detected,  the room
should be marked as occupied.

But we also find notable ambiguity in such design, caused by two factors. {\em First}, 
$\overline{\sigma_{aCSI}}$ depends on the target-anchor distance
and the motion pattern/strength. Yet the attacker has no knowledge of 
target behaviors or previous ground truth.   {\em Second}, short
physical distance to an anchor does not
always translate into being the same room. 

Next, we illustrate the problem of sensing ambiguity using real-world 
examples, including four basic cases with a single user and three complex cases with multiple users. 

\npara{Case 1: Target staying in a room.}
\yzedit{Figure~\ref{fig:csi_single}a} shows the traces of $\overline{\sigma_{aCSI}}$ for
three anchors: A and B in the left room, and C in the right room. The target user stays inside the left room
and moves around anchor A. In this case,
anchors B and C show no sign of targets nearby (very low  
$\overline{\sigma_{aCSI}}$) while 
anchor A has the largest $\overline{\sigma_{aCSI}}$ over time. 

\npara{Case 2: Target moving across rooms.} Following case 1, 
the target walks towards the room door (already open) at $t=12s$,
enters hallway at $t=18s$,  starts to open the right room door at
$t=24s$,  closes it and enters the room at $t=28s$.   In this case, anchor A's $\overline{\sigma_{aCSI}}$ drops as the target
moves away, followed by a short, minor increase due to the
opening/closing of the right room
door.   Anchor C has a short, minor increase in its
$\overline{\sigma_{aCSI}}$ due to the door
opening/closing, followed by a significant increase as the target
moves closer.  Interestingly, as the target transitions between the two rooms, we can
observe somewhat synchronized changes on anchor A and C (since  they
are triggered by the same event).  But from per-anchor
$\overline{\sigma_{aCSI}}$ traces,  a naive design will mark both rooms
as occupied.

\npara{Case 3: Sniffer blocked by external pedestrian.}
Pedestrians who move outside of the target area near the attack sniffer
could also create aCSI variations. Yet such movements (near the
common receiver) will create synchronized aCSI 
variations at all the transmitting anchors, regardless
of any human presence. Again a naive design will mark both rooms
as 
occupied.

\npara{Case 4: External users walking around the house.}
When pedestrians move away from the
sniffer, the impact on $\overline{\sigma_{aCSI}}$ 
is small even when they are close to the anchors 
(\yzedit{Figure~\ref{fig:csi_single}d}). This is because those
movements have little impact on the multi-path propagation  between
the anchors (inside the two rooms) and the sniffer.

\npara{Case 5: Multiple users moving in neighboring rooms.}
\yzedit{Figure~\ref{fig:csi_multi}a} shows an example where two targets are moving
in two different rooms, each with an anchor device. In this case, both
anchors (A and C) display large $\overline{\sigma_{aCSI}}$.

\npara{Case 6: Multiple users moving in distant rooms.} A user walks around in room A when another user 
sits down near an anchor in room C
(\yzedit{Figure~\ref{fig:csi_multi}b}).  We see that room A and C's anchors
are triggered, but not the one in room B.

\npara{Case 7: Anchors on both sides of a wall.} 
\yzedit{Figure~\ref{fig:csi_multi}c} shows that when the user moves near
anchor A, it triggers both anchor A and B (on the other side of
wall).  Here the simple design will mark both rooms as occupied
(since both anchors are triggered), creating a false positive. 

\subsection{Design Concepts}

Our analysis shows that instantaneous $\overline{\sigma_{aCSI}}$ observed
at each individual anchor is insufficient to detect and
localize user
motion.  We propose to overcome sensing ambiguity by analyzing the value
and pattern of $\overline{\sigma_{aCSI}}$ across both time and 
anchors. The end result is  a robust $\overline{\sigma_{aCSI}}$ model that
links each human motion with signal dynamics of anchors in its actual
room.  Next, we outline the signal analysis process in two
steps: 1) {\em detecting human motion} and 2) {\em mapping each
  motion to a room}.  The detailed procedures are described later in \S\ref{sec:locate_mobile}.

\para{Detecting human motion.}  If the number of detected
anchors per room is reasonable\footnote{Home/office WiFi devices
 naturally 
  spread out in a room~\cite{iotlayout1,iotlayout2}. One can assume
  3-4 devices in a room of common size of $25 m^2$.}, 
any
user movement should ``trigger'' at least one anchor in the scene. 
But {\em how do we determine threshold $\sigma_{p}(x)$ necessary to trigger an
  anchor $x$}?  This is not easy, because the adversarial has no ground
truth on target presence. Also the threshold should be anchor-specific and
could also vary over time. 

Leveraging a common observation where a user will not 
  stay and move in a single room forever, we  
propose to derive $\sigma_{p}(x)$ by finding
``outliers.''  Assuming for anchor $x$ the sniffer can measure
$\overline{\sigma_{aCSI}}(x)$ over a long period of time ({\em
  e.g.\/}, hours or even days),   it is reasonable to assume that $x$ is mostly
not triggered.   Thus the sniffer can apply outlier detection methods
like MAD~\cite{mad,mad_asym} to derive $\sigma_{p}(x)$ and 
adapt it
over time.

\para{Mapping Each Motion to a Room}  When multiple anchors in more than one
room are triggered, the sniffer needs to decide whether they are triggered by users in one
 room (one source) or users in multiple rooms (multiple sources).  This is because a target's movement
 could trigger anchors in neighboring rooms (case 7), and the same
 holds when multiple users move in two rooms (case 5 and 6). The
 sniffer needs to distinguish between them and determine the set of
 rooms that are actually occupied. 

Again we leverage a common observation: human 
movements in different rooms are generally asynchronous, thus anchors
triggered by separate sources will display different temporal
patterns in $\overline{\sigma_{aCSI}}$ (case 5 and 6).  But when a single source triggers anchors in neighboring rooms (case 7),
these anchors' $\overline{\sigma_{aCSI}}$ will share a similar
pattern.  By computing the correlation of normalized
$\overline{\sigma_{aCSI}}$ time series across anchors, we
can determine whether they are triggered by sources in one
room {\em i.e.\/} positively
correlated.  For example, the correlation 
between the two triggered anchors are -0.07, -0.03, and 0.32, in case
5, 6, and 7, respectively, and 0.23 during the door opening in
case 2. 

Our attack can also use the floor plan (or room transition
probabilities) to fine-tune the detection result (similar to~\cite{dina-ubicomp18}).  For example, a user cannot ``fly'' from
one room to another when the rooms are widely
separated. If the sniffer observes two anchors in two widely
separated rooms being triggered sequentially with little or no gap, it will
report two users, one in each room, rather than a single user moving
across rooms.  

For other cases, our attack 
conservatively treats the rooms with at least one anchor triggered as
occupied.

\section{Attack Design}
\label{sec:design}
\begin{figure}[t]
\centering
  \includegraphics[width=0.4\textwidth]{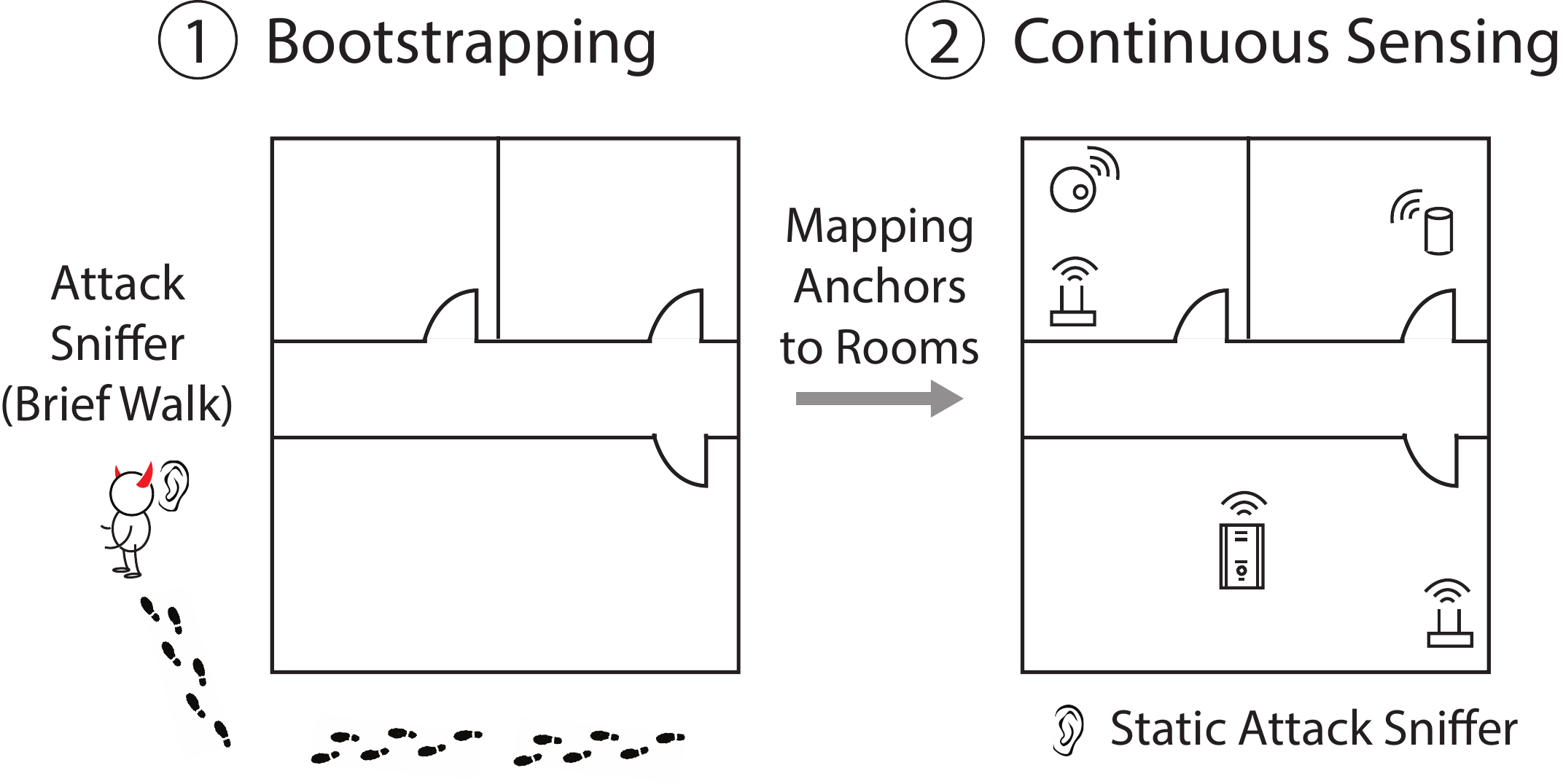}
  \caption{Our attack process includes a
    bootstrapping phase and a continuous sensing phase.}
  \label{fig:attackprocess}
\end{figure}

After presenting the key concepts, we now
present the attack design in detail. 
As shown in Figure~\ref{fig:attackprocess}, the attack
includes two phases: (1) identify and locate anchors
during ``bootstrapping,'' and (2) deploy the sniffer and perform ``continuous
human sensing.''

\para{(1) Bootstrapping.}  The attacker first needs to identify and
locate the anchors in the target area.  The unique feature of our
motion detection is that is does not require precise location of anchors,  only
their individual room.  In our attack, this is achieved by the
attacker performing
a brief passive WiFi
measurement (using the sniffer) while walking outside the target
property. Similar to the evil maid
attack~\cite{evilmaid}, the walking measurements are only necessary 
during initial bootstrapping.

Before feeding the collected measurements into a device
localization algorithm,  our attack introduces a novel {\em data
  sifting} procedure to identify the right measurement instances for
anchor localization.  As a result, the attacker can localize anchors
down to their individual rooms using limited and often noisy signal
measurements\footnote{Because the attacker has little control 
on the available walking path
and the propagation environment,  the signal measurements will
inevitably contain
bias, noise and human errors.}. 

\para{(2) Continuous human sensing.} After locating a list of anchors, the
attacker hides the same sniffer at a fixed 
  location outside the target area. The sniffer continuously monitors ambient
  WiFi signals, and uses them to locate and
  track human presence and movements inside.  The sniffer also monitors each
  detected anchor, and any relocation of an anchor will trigger its removal
  from the anchor list, and possibly another {\em bootstrapping} phase to
  (re)locate the anchors.

\shepherd{Our proposed attack process is fully automated, and does not require 
any operations by the adversary, beyond the initial bootstrapping which
involves a walk around the property to collect signal measurements. Note
that this walking measurement could also be achieved by a robot/drone.}

\subsection{Continuous Human Sensing}
\label{sec:locate_mobile}
In this phase, the sniffer will continuously collect
$\overline{\sigma_{aCSI}}$ for each anchor and analyze them to
detect, locate human presence to their individual rooms.

\para{Detecting the presence of human motion.}   For each 
anchor $x$, when 
$\overline{\sigma_{aCSI}}(x)>\sigma_{p}(x)$,  the sniffer declares the presence of motion near
$x$, or ``anchor $x$ is triggered.''  To compute  $\sigma_{p}(x)$, the
sniffer applies median
absolute deviation (MAD)~\cite{mad,mad_asym} on observed 
$\overline{\sigma_{aCSI}}(x)$ over time. 
Assuming ``untriggered''  $\overline{\sigma_{aCSI}}(x)$ follows a
Gaussian distribution, we have 
\begin{equation} 
  \sigma_{p}(x)=\lambda\cdot \mbox{MAD}(Z)+\mbox{median}(Z)
\end{equation}
where $\lambda$ is the conservativeness factor and $Z$ is the
long-term observation of  $\overline{\sigma_{aCSI}}(x)$.
\shepherd{The choice of $\lambda$ will affect the motion detection
  rate and false alarm rate.  For our attack, we set $\lambda=11$ so
   the ideal detection rate per anchor is high. }

\para{Assigning target(s) to rooms.}  When 
any anchor(s) get triggered,  the sniffer analyzes their temporal $\overline{\sigma_{aCSI}}$ traces to
determine the set of rooms that are actually occupied.

 (i) If all the
triggered anchors are in the same room, then the room is declared as
occupied. Exit. 

(ii) If most of the anchors are triggered, and their
$\overline{\sigma_{aCSI}}$ time series are (consistently) positively correlated,  then
the sniffer is blocked by an external pedestrian next to the sniffer,
and the sensing output is 
``uncertain.'' Exit. 

(iii)  Now consider all the triggered anchors. Start from the
triggered  anchor $x$ with the highest
$\overline{\sigma_{aCSI}}$.  Mark $x$ as ``checked'' and its room
as occupied. Compute pair-wise correlation between $x$
and any triggered anchor ($y$) in neighboring rooms.  If $x$ and $y$
are highly positively correlated, mark $y$ as checked. Repeat until
all the triggered anchors are ``checked''.

\para{Tracking targets.}  After detecting a set of motion events, the
sniffer can potentially 
combine them with room transition probabilities built from the floor plan to 
estimate user trajectories.  
\shepherd{For example, the sniffer can track a security guard's patrol
  route from a sequence of detected motion events. 

It should be noted that  while our attack can detect whether a room is
occupied or not, it cannot identify an individual out of a group of
users in the same room.  Thus accurate per-person tracking is {\bf only} feasible
when the number of users is small. }

\para{Monitoring anchor status.}  The sniffer also monitors each
(static) anchor's RSS (see \S\ref{sec:locate_static}). Upon detecting a 
considerable RSS change for an
anchor (\shepherd{which indicates potential relocation}), the attacker either removes it from the anchor list or run
bootstrapping to relocate anchors and recompute its $\sigma_{p}$. 

\para{Impact of sniffer placement.} The sniffer should be placed where
it can capture aCSI signals from the detected anchors, while avoiding
being too close to the anchors or at busy places with pedestrians frequently
passing by.  While one could further optimize the sniffer location,  our current
design randomly chooses a location that can capture signals from all
the anchors.

\subsection{Bootstrapping: Locating Anchors}
\label{sec:locate_static}
During bootstrapping, the attacker uses the passive sniffer to
identify and localize {\em static} anchors inside the target property.  
There are many device localization proposals, but since the sniffer 
stays passive and only has
a single antenna,  we choose to use RSS-based method~\cite{liqun14mobicom,zhijing18hotmobile}.  In this case, with a brief walk outside of the target's home/office,  the
adversary uses the sniffer to measure RSS of potential anchors along the
trajectory. These spatial RSS values and the trajectory (recorded by
the smartphone's IMU sensors) are fed into a log distance path loss
model~\cite{pathlossmodel} to estimate the transmitter location.
Each 
transmitter located inside the target scene area is added to the anchor
list.

\para{Why RSS but not aCSI?} The localization uses RSS 
rather than aCSI, fCSI or AoA~\cite{multipathtriang18mobisys,
  mostofi18ipsn} because of 
two 
reasons.
  {\em First}, our attacker sniffer only has one antenna, and cannot 
  estimate fCSI accurately due to lack of synchronization with
  the transmitter.   Recent work~\cite{mostofi18ipsn} estimates AoA
  from aCSI, but only if the sniffer has an antenna array and is in 
  complete LoS to the targets, {\em i.e.\/} no wall. 
 {\em Second}, as shown in \S\ref{sec:locate_mobile}, aCSI is sensitive to nearby target
movements. As the adversary has no knowledge of the
target status during bootstrapping, it cannot rely on aCSI for localization.  
In comparison, RSS is much more robust against 
target movements, thus a reliable input for localization under the
adversarial scenario. 

\shepherd{
  \para{Identifying static anchors.} 
}
RSS of a static transmitter,
when captured by a static sniffer, should stay relatively
stable, while those of moving ones will fluctuate over time.
Thus before running spatial RSS measurements, the attacker will
keep the sniffer static and measure the per-device RSS standard
deviation ($\sigma_{RSS}$) for a period of time
({\em e.g.\/} 60s).  Devices with large $\sigma_{RSS}$ ($>$2.7dB in
our work)  are not used as
anchors. This is repeated during
the continuous sensing phase (see 
\S\ref{sec:locate_mobile}) to detect relocation of any anchor device.
A complementary method is to infer the device type (and brand name) from the Organizational Unique Identifier (OUI) field of
  the MAC  address~\cite{privacy14iccst} and ignore moveable
  devices like smartphones, wearables, laptops, and camera robots.

\para{Finding high-quality RSS measurements.}  The
localization accuracy depends heavily on the ``quality'' of RSS
measurements.  Instead of searching for a new localization design, we
apply a statistical data sifting algorithm to identify proper RSS
data samples as input to the localization algorithm.

The attacker can filter out ``bad'' measurements using
de-noising methods, {\em e.g.}, Kalman
filter~\cite{kalmanfilter}, wavelet filter~\cite{wavelet16mobihoc} and
feature
clustering~\cite{zhijing18hotmobile}. We find that these are
insufficient under our attack scenarios because the propagation
environment is highly complex and unknown to the adversary, making it hard to distinguish between noise and natural propagation
effect.  Similarly, features used by~\cite{zhijing18hotmobile} to identify bad
measurement rounds are too coarse-grained to
effectively control localization
accuracy.  In fact, our experiments in \S\ref{sec:eval} show that
with~\cite{zhijing18hotmobile},  $>50\%$ of the good measurement
rounds it identifies all locate the device to a wrong room.

Instead, we propose {\em consistency-based data sifting} to
identify proper data samples that will be
used for model fitting. 
Our
hypothesis is that, by the law of large numbers~\cite{lawlargenumber},   
{\em consistent} fitting results from many random samplings of RSS measurements, if exist, can reveal true signal
propagation behaviors and produce high-fidelity localization results.

\begin{figure}[t!]
 \centering
 \includegraphics[width=0.38\textwidth]{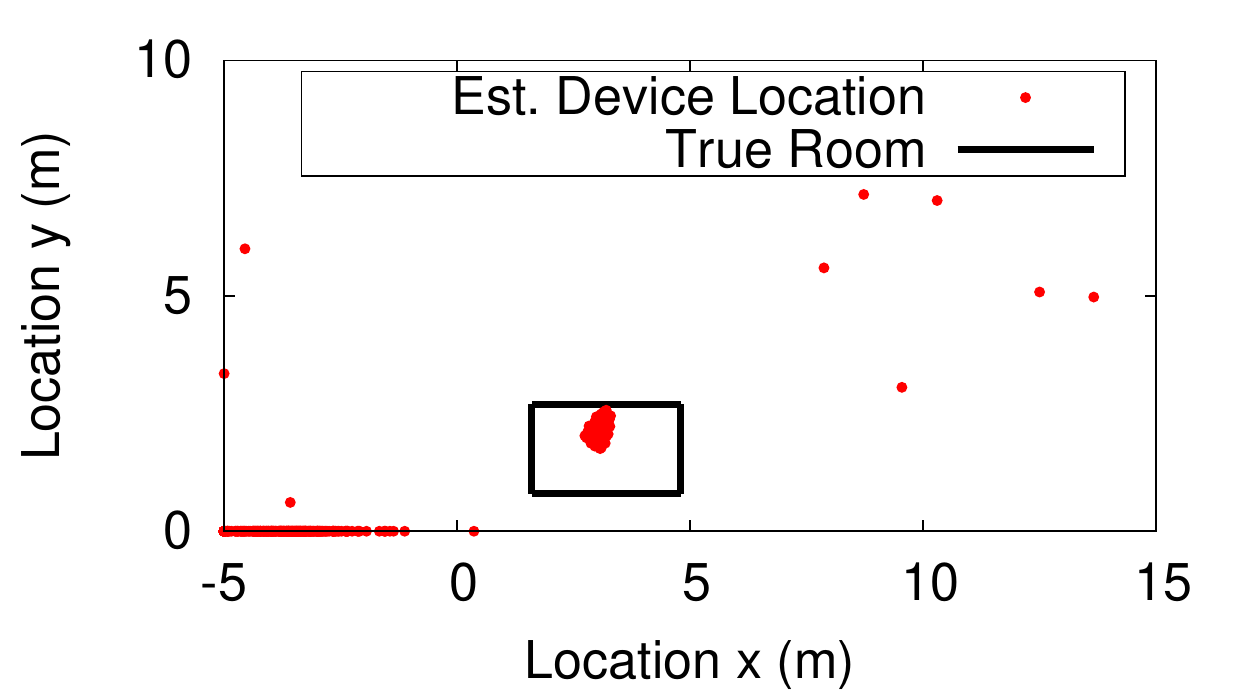}
\caption{Improving accuracy of anchor localization using our proposed
  consistency-based data sifting.  Each red dot is the anchor location
  estimated from a Monte Carlo sample of RSS measurements. The
  rectangle marks the actual room the anchor resides. In this example,
  a dominant cluster is present and is used to estimate the final
  anchor location.}
 \label{fig:cluster}
\end{figure}

Given a round of measurements $\mathbb{R}$, we apply the Monte Carlo
  method~\cite{montecarlo} to randomly sample a subset (80\%) of $\mathbb{R}$ as the input to the model fitting.  This is
repeated by $N=1000$ times, producing $N$ localization
results.  We can find natural clusters among these $N$ results 
from their locations and fitting mean square errors (MSE).  
If they form many small
clusters with different room placements, then $\mathbb{R}$ is  
inconsistent and cannot be used for localization.
If a dominant cluster exists and its averaged MSE is
less than those of the other clusters, then $\mathbb{R}$ can be used for
localization.  An example is shown in Figure~\ref{fig:cluster}, which produces a single,
dominant cluster, while the rest are widely
scattered.  When such a dominant cluster is present, we can estimate the anchor
room location by aggregating the location data points of the cluster.
In the example of Figure~\ref{fig:cluster}, all the data
points are located in the top center of a single room.  When the data
points belong to different rooms, we choose the room with the most
data points.

When multiple rounds of RSS measurements are available, the attacker can apply consistency
check -- if a localization
  result is consistent across multiple rounds, then it is a confident
  result.  Across our experiments,  we find that consistency check across 4 rounds of measurements is
  sufficient to achieve a room placement accuracy of 92.6\%.

\para{Floor-level signal isolation.}
\label{subsec:floordetection}
When the target property has multiple floors, the attacker needs to
localize wireless anchors to a particular floor during bootstrapping. 
This is easily achieved using coarse angle of arrival (AoA) estimates
captured by the smartphone with a simple cone cover to focus signals from a
particular AoA. The received RSS from each anchor can be combined with the
phone angle (via the built-in gyroscope) to localize each anchor to a floor.

\section{Smartphone Implementation}
\label{sec:imple}

We prototype our attacker system using a commodity smartphone as the sniffer.
We implement the bootstrapping and continuous sensing modules each as
an Android app, and deploy and experiment using two
versions 
of Android phones, Nexus 5 and Nexus 6. Both smartphones are equipped with the Broadcom WiFi
chipset. For spatial RSS measurements (during bootstrapping), we use the built-in
IMU sensors (accelerometer and gyroscope) to detect user strides
and build trajectory.  \shepherd{The key system parameters are listed in
Table~\ref{tbl:parameters}. }

\para{Enabling continuous, passive sniffing of aCSI.}  Previously, aCSI can only be
captured when the receiver actively communicates with the target 
transmitter~\cite{Halperin_csitool}.  Recent
work~\cite{shadowwifi18mobisys} produces a firmware (Nexmon) that
enables passive\footnote{Passive sniffing means
  that the sniffer does not need to communicate with the target
  transmitter, thus remains completely undetectable.} sniffing, but only on a single customized transmitter at very low
rates.  

For our attack, we made a series of changes to the Nexmon firmware, so
that the sniffer can run continuous passive sniffing and capture aCSI
from multiple commodity WiFi devices simultaneously.  In particular,
we made changes to hardware buffer management to resolve the issue of
buffer overflow facing the original  Nexmon firmware. 

One remaining artifact is that the firmware only reports aCSI 
at a limited speed, up to 8--11 packets per second. To save energy,
we subsample sniffed
packets based on this rate limit.  Despite this artifact, 
our prototype  sniffer is able to capture
sufficient aCSI samples to successfully launch the attack.

\para{Computation and energy cost.} One strength of our attack is its
simplicity.  For our current smartphone prototype, the bootstrapping
app runs 1000 rounds of Monte Carlo sampling and model fitting, which
finishes in less than 25s per anchor. It takes less than 1s to compute
average aCSI standard deviation. 
The app consumes 4.18 watts
(bootstrapping) and 2.1 watts (continuous sensing), respectively. 
For Nexus 5 (with a built-in 2300mAh battery),  this enables 4.1 hours of continuous
sensing. 
Currently our app does not optimize for energy efficiency, which
could be improved to extend sensing duration. 

\begin{table}[t]
  \centering
  \resizebox{0.41\textwidth}{!}{
\begin{tabular}{|l|l|}
\hline
\textbf{Parameters}                            & \textbf{Value}           \\ \hline
MAD conservative factor $\lambda$              & 11                       \\ \hline
Threshold of $\sigma_{RSS}$ for static anchors & 2.7                      \\ \hline
Ratio of Monte Carlo sampling size             & 80\%                     \\ \hline
Number of Monte Carlo sampling rounds ($N$)    & 1000                     \\ \hline
\end{tabular}
}
\vspace{0.04in}
\caption{\shepherd{Attack parameters used in our experiments.}}
\label{tbl:parameters}
\end{table}

\begin{table}[t]
\centering
\resizebox{0.32\textwidth}{!}{
\begin{tabular}{|c|c|c|c|}
\hline
Sniffer & Test  & \# of & Mean Room   \\
Path    & Scene & Rooms & Size ($m^2$)  \\\hline 
\multirow{9}{*}{\begin{tabular}[c]{@{}c@{}}Indoor\\Hallway\end{tabular}} 
 & 1 & 6 & 14.19 \\ \cline{2-4} 
 & 2 & 7 & 14.60    \\ \cline{2-4} 
 & 3 & 8 & 13.65  \\ \cline{2-4} 
 & 4 & 3 & 14.50   \\ \cline{2-4} 
 & 5 & 3 & 9.51  \\ \cline{2-4} 
 & 6 & 6 & 14.21   \\ \cline{2-4} 
 & 7 & 5 & 16.75   \\ \cline{2-4} 
 & 8 & 4 & 44.39  \\ \cline{2-4} 
 & 9 & 2 & 69.83   \\ \hline
\multirow{2}{*}{\begin{tabular}[c]{@{}c@{}}Outdoor\\Sidewalk\end{tabular}} 
 & 10 & 2 & 47.20   \\ \cline{2-4} 
 & 11 & 4 & 12.99   \\ \hline
\end{tabular}
}
\vspace{0.04in}
\caption{Test scene configuration.}
\label{tbl:attack_environments}
\end{table}

\section{Evaluation}
\label{sec:eval}

We evaluate our attack using experiments in typical office buildings and
apartments. 
We describe our experiment setup and test scenes, 
present our evaluation on individual attack phases (bootstrapping and
continuous sensing), followed by an end-to-end attack evaluation.

\begin{figure*}[t!]
\centering
\mbox{
\subfigure[Scene 4]{
  \includegraphics[width=0.30\textwidth]{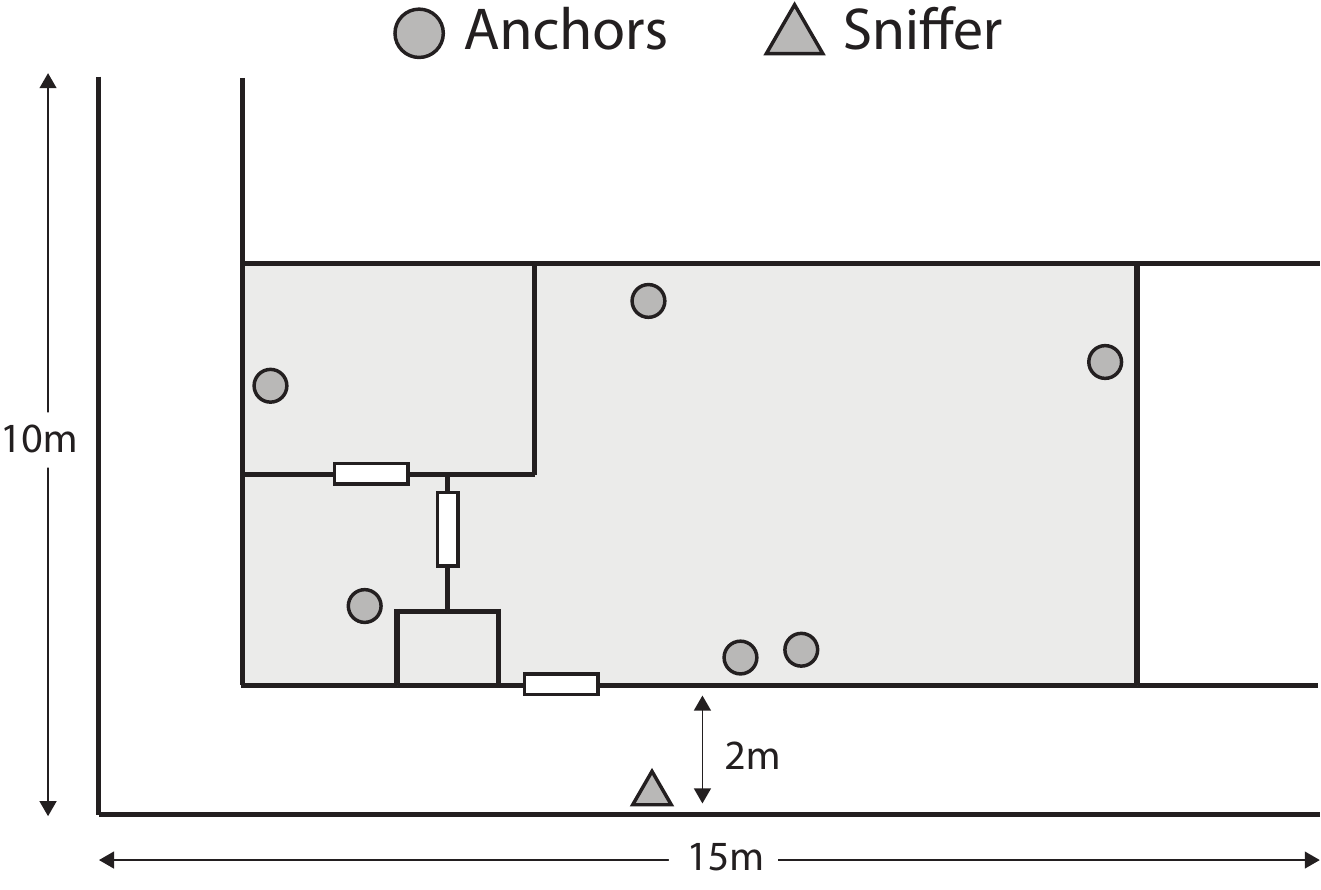}
}
\subfigure[Scene 6]{
  \includegraphics[width=0.24\textwidth]{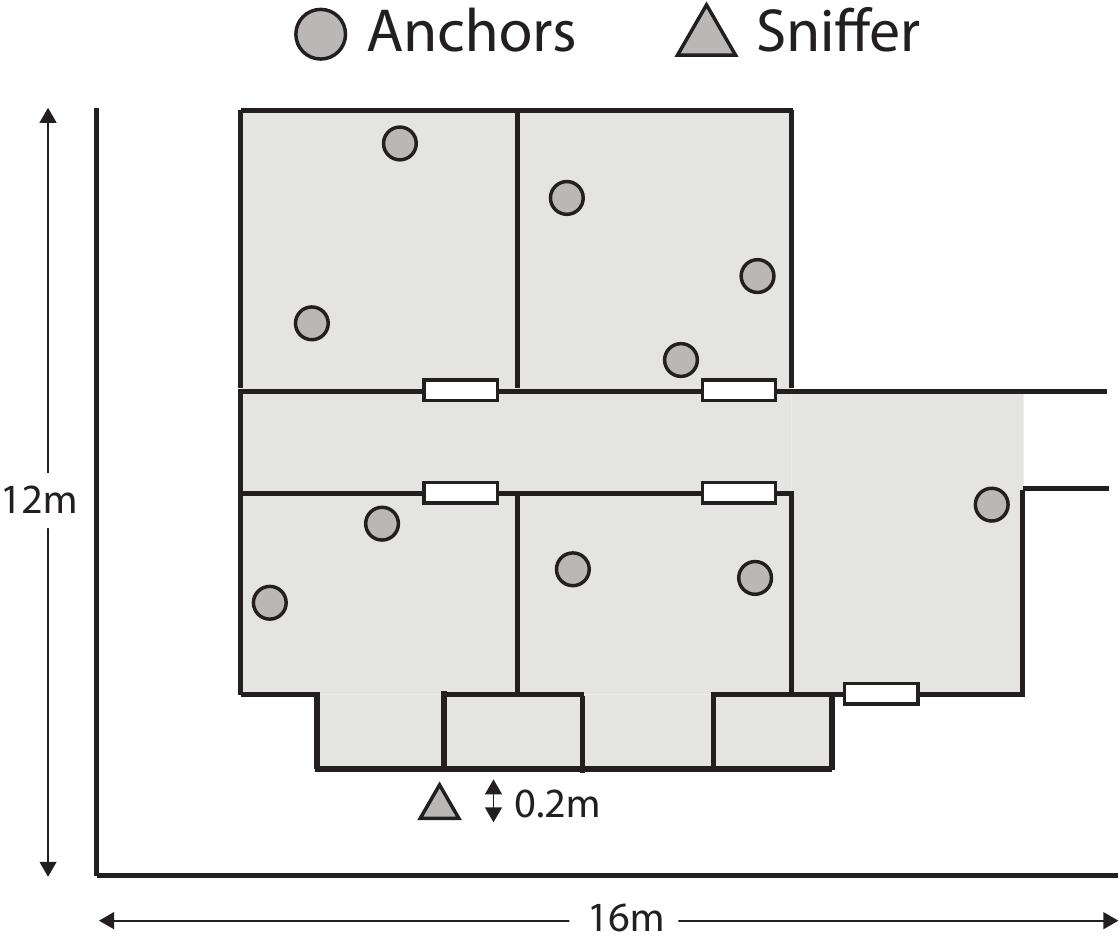}
}
\subfigure[Scene 8]{
  \includegraphics[width=0.32\textwidth]{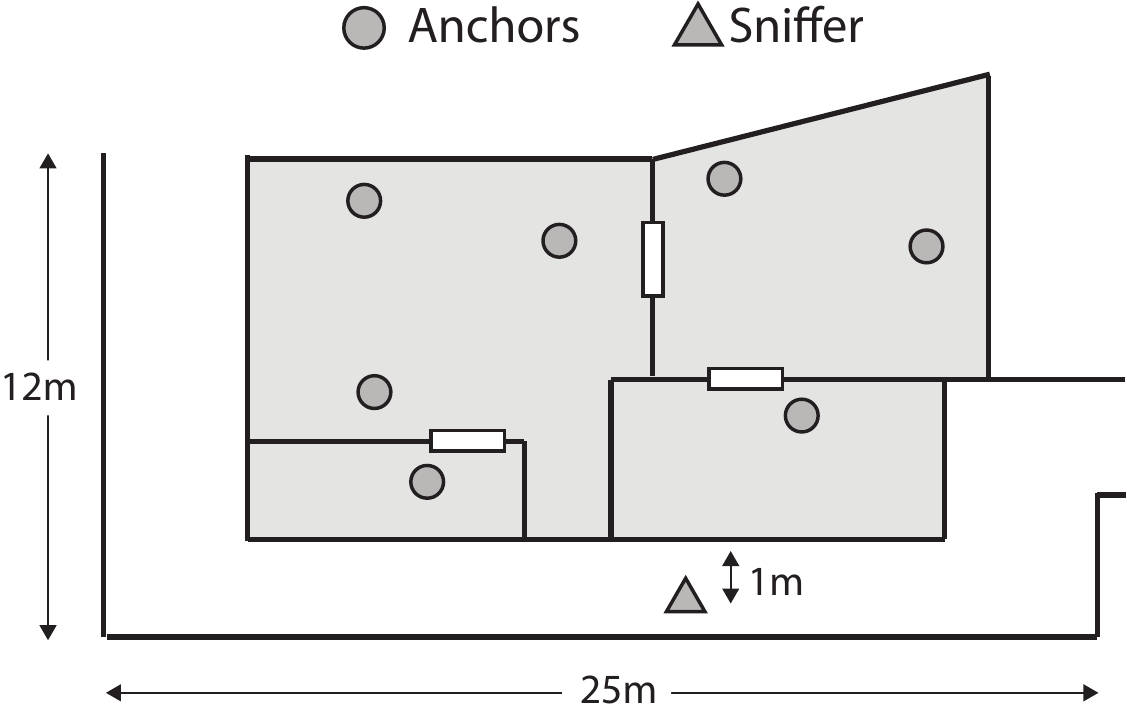}
}
}
\caption{Sample test scene floorplans, derived from the real estate websites
  or emergency exit maps, where shaded regions are the target
  property.   We also show an instance of anchor
  placements where $\bigcirc$s are the anchor devices,
 and $\bigtriangleup$ is the
static attack sniffer.}
\label{fig:attack_floorplan_example}
\end{figure*}

\begin{table*}[t]
\centering
\resizebox{0.9\textwidth}{!}{
\begin{tabular}{|c|c|c|c|c|}
\hline
& \multirow{2}{*}{Device Type} &  \multirow{2}{*}{Exact Product} & \yzedit{Mean} Packet Per
  & \yzedit{Mean} Packet Per \\ 
& & & Second (pps), Idle & Second, Active \\ \hline
\multirow{5}{*}{Static}
& Cameras (w/o Motion Detection) & AHD Security Camera &  N/A & 124 \\ \cline{2-5}
& Cameras (w/ Motion Detection) & Amcrest/Xiaomi IP Camera & $\ge$0.5 & 108 \\ \cline{2-5}
& Home Voice Assistance & Amazon Echo, Google Home & 2 &  16 \\ \cline{2-5}
& Smart TV (\& Sticks) & Chromecast, Apple TV, Roku & 6.64 & 200 \\ \cline{2-5}
& Smart Switches & LifeSmart Plug & $\ge$2.44 & $\ge$3.33 \\ \cline{2-5}
& WiFi Router & Xiaomi/Cisco/Asus Routers & 28.6 & 257 \\  \hline
\multirow{3}{*}{Mobile}
& Surveillance Robot & iPATROL Riley Robot Camera & N/A & 124 \\ \cline{2-5}
& Smartphones & Samsung/Google/Apple Phones & $\ge$0.5 & $\ge$6  \\  \cline{2-5}
\hline
\end{tabular}
}
\vspace{0.04in}
\caption{Summary of WiFi devices used in our experiments. Note that our attack
will detect and recognize static anchors and only use them to
detect/localize human motion.}
\label{tbl:devices_summary}
\end{table*}

\vspace{-0.05in}
\subsection{Experiment Setup}
\label{sec:setup}

 We experiment at 
11 typical offices and apartments that are
accessible to us.  The owners of each test volunteered for our
experiments.  The test scenes are of different sizes and configurations, and
have different wall materials.  The walking path available to the
adversary also differs across experiments, from indoor corridors outside
the apartment to outdoor
pathways. Table~\ref{tbl:attack_environments} lists the test scene
configuration while Figure~\ref{fig:attack_floorplan_example} shows 
floor plan examples \shepherd{derived from publicly available
  data.  Across all experiments, attack parameters remain
unchanged (as listed in Table~\ref{tbl:parameters}). }

Inside each test scene, we either reuse existing WiFi devices or deploy our
own WiFi devices to emulate smart homes and offices. 
We use popular commodity products for smart
offices and homes,  {\em e.g.}, wireless
security cameras, voice assistants, WiFi routers, and smart
switches. In total, we have 31 WiFi devices, including 6
security cameras. These devices are naturally placed at locations where they are designed
to be:  security cameras at room corners, smart switches on the wall
outlets,  and WiFi routers in the center of the room
for coverage. Our experiments use the 2.4GHz WiFi band due to its dominant coverage.
  We also test 5GHz WiFi and do not observe notable difference
  except its shorter coverage.

Table~\ref{tbl:devices_summary} summarizes these devices
and their traffic patterns during idle and active periods. 
The packet rate varies from 0.5 packet per second (pps) to
more than 100 pps.   Even when idle, they still periodically transmit
packets.  It should be noted that to prevent attackers from inferring
user presence by simply counting the packet rate of a device (if an Amazon
Echo is sending more packets, it means that a human user is around),
devices like home voice assistants, smart TVs, and motion-triggered
cameras will
need to send cover traffic when in idle state and the
corresponding idle packet rate will be much higher than the listed
number.

\para{Bootstrapping.} To benchmark our bootstrapping design,  we
collect, for each test scene, 50 walking measurements,
each of 25--50 meters in length and 0.5--2 minutes in time.
We also change anchor placements and repeat the experiments.  In
total, we collect more than 3000 RSS measurement traces, with more
than 121,000 location-RSS tuples.

\para{Continuous sensing.} We place a static sniffer behind plants or at the corners (on the ground) outside of
the target building within $2m$ to the building wall.
We ask volunteers 
to carry out normal activities in each test scene and
collect more than 41hrs of aCSI entries (7.8hrs of human presence, labeled).
The volunteers
are aware of the attack goals but not the techniques.

\subsection{Evaluation of Continuous Human Sensing}
\label{subsec:sensingeval}

We start from evaluating the {\em continuous sensing} component of our
attack.   Here we assume that the attacker knows the actual room where
each anchor resides. By default, the attacker only uses anchors whose packet rate
$\geq$ 11pps.

\para{Performance metrics.} Our goal is to evaluate whether the
continuous sensing component is able to correctly detect user
presence/motion in each room.  We divide time into 5s slots, and
run continuous sensing to 
estimate room occupancy \shepherd{in each slot} based on aCSI variance values.  We compare
these estimates to ground truth values, and compute the detection
rate and false positive rate as follows.

\begin{packed_itemize}
\item {\em Detection rate} (DR) measures the probability of the attack reporting a room
as being occupied when it is actually occupied, \shepherd{across all the slots.}

\item {\em False positive rate} (FP) measures the probability of a
  room not being occupied
when our attack reports that it is being occupied. 
\end{packed_itemize} 
Under our adversarial scenario, having a high detection rate is 
more important since the attacker does not want to miss the presence of any targets.

\para{Human sensing accuracy.} 
Table~\ref{table:userdetection} lists the detection rate and false
positive rate when we vary
the number of anchors per room.  We see that the detection rate
scales with the number of anchors per room, reaching 86.8\%, 95.03\% 
99.85\%, and 99.988\% with 1, 2, 3, and 4 anchors per room, respectively.  This
trend is as expected since having more anchors increases the chance
that a user movement triggers at least one anchor. Furthermore, the
false positive rate is low ($<$3\%) with a single anchor per
room and increases slightly to 6.9\% if the attacker wants to leverage
all 4 anchors.  Across our experiments, the false
positives mainly come from the impulse noises in aCSI reported by
the firmware.  Thus having more anchors will lead to more false
positives.

\begin{table}
    \resizebox{\columnwidth}{!}{
\begin{tabular}{cc|c|c|c|c|}
\cline{3-6}
                &           & \multicolumn{4}{c|}{\# of WiFi Devices Per Room} \\ \cline{3-6}
                                                &           & 1
                                                & 2          & 3
                                                & 4         \\ \hline
  \multicolumn{1}{|c|}{\multirow{2}{*}{Ours}} & DR  & 86.824\%    & 95.034\%    & 99.854\%    & 99.988\%   \\ \cline{2-6}
  \multicolumn{1}{|c|}{}                          & FP & 2.927\%    & 4.082\%  &  5.305\%   &  6.935\% \\ \hline \hline 
  
\multicolumn{1}{|c|}{\multirow{2}{*}{LiFS }} & DR    & 20.536\%    & 37.040\%    & 50.262\%    & 60.821\%   \\ \cline{2-6}

  \multicolumn{1}{|c|}{}                           & FP &  4.622\%  & 4.961\% & 5.395\% &   5.886\% \\ \hline \hline 
\multicolumn{1}{|c|}{\multirow{2}{*}{LiFS}} & DR    & 43.568\%    & 68.315\%    & 82.289\%    & 90.149\%   \\ \cline{2-6}
  \multicolumn{1}{|c|}{(unrealistic)}                          & FP&4.622\%& 5.364\%&6.443\%&7.644\%\\ \hline 
\end{tabular}
}
\vspace{0.04in}
\caption{Detection rate (DR) and False positive rate (FP) of continuously
  human sensing, assuming accurate room placement of anchors.  We
  compare our design to the state-of-art human sensing system (LiFS).}
\label{table:userdetection}
\vspace{-0.3in}
\end{table}

We also compare our system to the current state of the art of passive
human sensing (LiFS~\cite{lifs16mobicom}). For fair comparison, we
add \yzedit{wavelet denoising} to LiFS, confirming that it improves
the sensing performance.   Since LiFS requires each
anchor's precise physical location in the room (which is not available to our
attacker), we first use the room center as the input to LiFS, mapping to
1--2m localization error. LiFS also requires knowledge of the aCSI
value when no user is present, which we use the same MAD based method
to estimate.   Results in Table~\ref{table:userdetection} show that
even with four anchors in the room, LiFS can only achieve a detection rate of 
60.82\%. Here the miss-detection happens when LiFS locates the human
presence to a wrong room.   We also run another version
of LiFS that is unrealistic under our attack scenario, where each
anchor's physical location error is random but bounded by 50cm (and without any room
placement error).   In this case,
its detection rate improves, but is still far
from our attack, especially with smaller number of anchors per room.

\begin{figure}[t]
  \begin{minipage}{0.45\textwidth}
        \centering
        \includegraphics[width=0.8\textwidth]{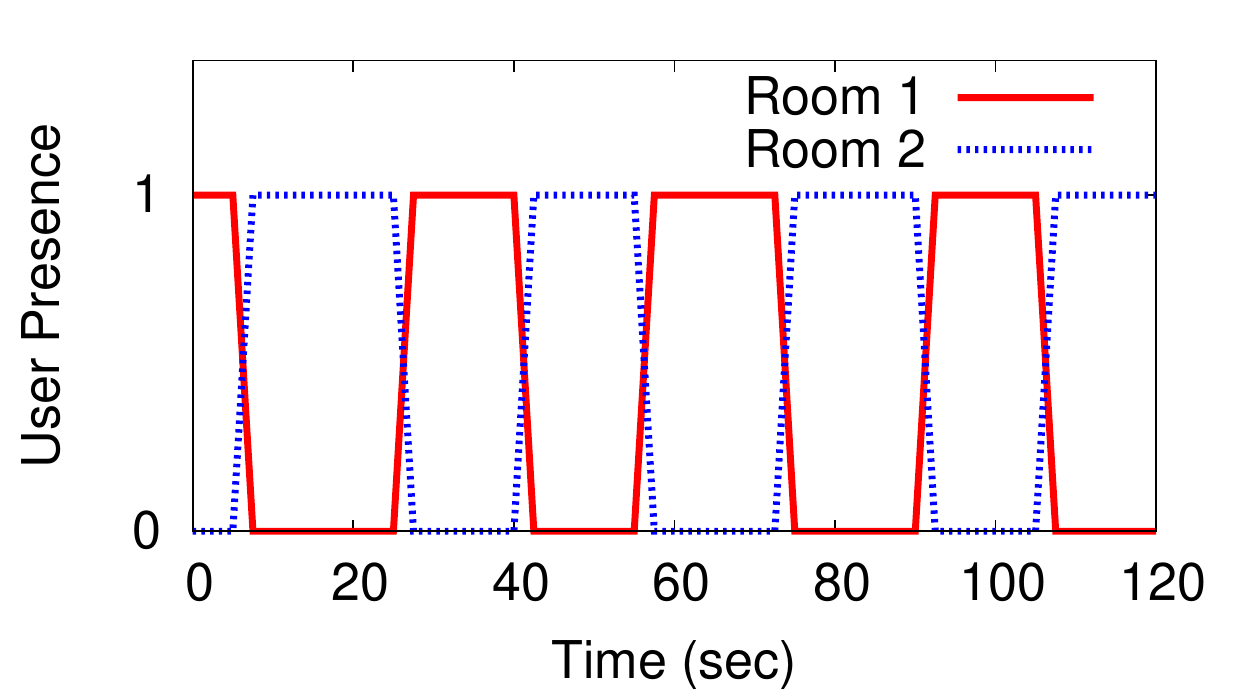}
        \vspace{-0.1in}
        \caption{The attack sniffer can track fast user motion between rooms.}
        \label{fig:csi_impact_periodic}
    \end{minipage}
   \hfill
    \begin{minipage}{0.45\textwidth}
        \centering
        \includegraphics[width=0.8\textwidth]{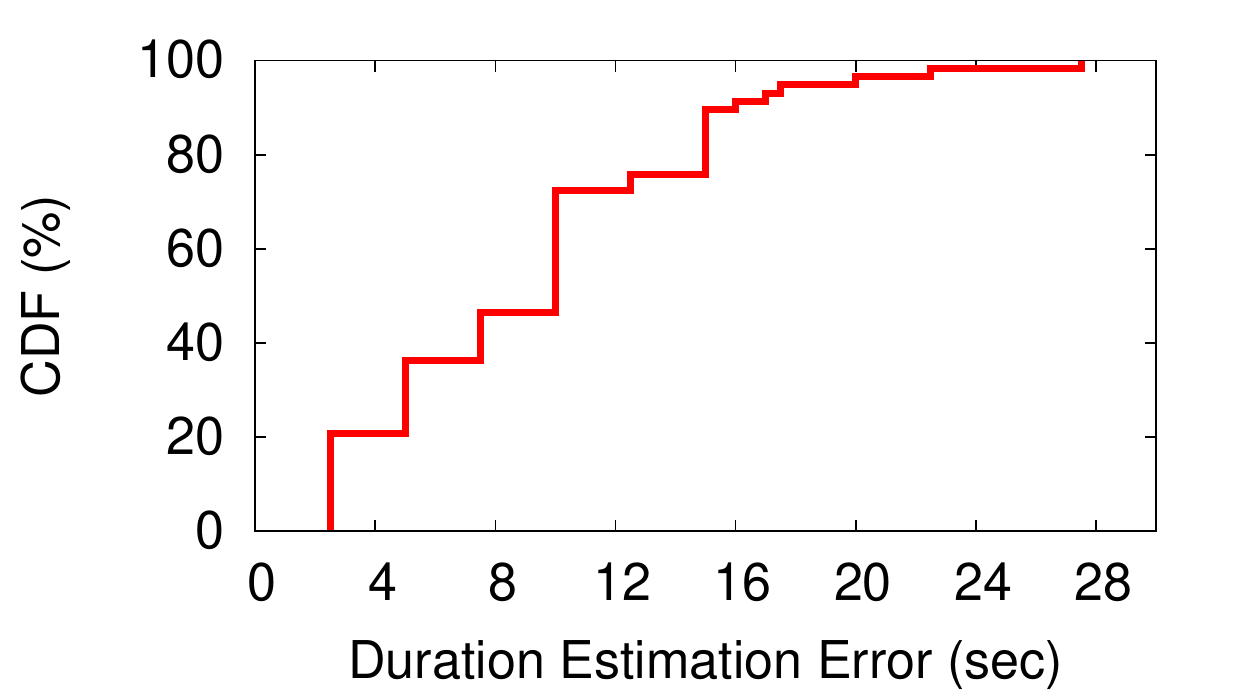}
        \vspace{-0.1in}
        \caption{Error in motion duration estimation is small.}
        \label{fig:estduration}
      \end{minipage}
  \vspace{-0.2in}
    \end{figure}

\para{Tracking responsiveness.} We also examine whether our
attack is able to track human movements in time.  We start from an
example scenario where a user moves
 back and forth between two connecting rooms, {\em i.e.\/} she walks
 in one direction for 18s, turns around and walks in the opposite
 direction, and repeats. Figure~\ref{fig:csi_impact_periodic} shows the
detected user occupancy of the two rooms (each with two anchors).  We
see that our detection is highly responsive to rapid human
movements.

We also consider all the aCSI traces collected across our test scenes and examine the duration of individual
movement events estimated by the attacker. We compare these estimations to the
ground truth. Figure~\ref{fig:estduration} plots the \shepherd{cumulative distribution function (CDF)} of the duration
estimation error, where for 80\% of the cases, the error is
less than 16 seconds.

\para{Impact of anchor packet rate.} So far, our results assume that
anchors send packets at no less than 11pps\footnote{ As discussed in
\S\ref{sec:imple}, the sniffer's firmware reports CSI in an equivalent packet rate of
8--11pps.}. 
To study the impact of anchor packet
rate,  we take the aCSI
traces of WiFi security cameras (w/o motion detection) and sub-sample them to
produce desired packet rates.
Our experiments show that for a single anchor per room, the detection rate is
86.824\% at its full
rate (an equivalent aCSI rate of 11pps), and 
reduces to 85.49\% at 2pps, and 69.29\% at 1pps. 
The false positive rate remains $<$5\%. This means that each low-rate
anchor can still ``help'' the attacker
identify and locate targets. For a room with multiple low-rate
anchors, the attacker will take the {\em union}
of their detection results.

\shepherd{
\para{Impact of interference.} 
During all experiments, we observe minimal impact on attack performance from interference 
by other WiFi transmissions out of our control or access.
In the presence of strong interference, anchor packet rates will drop
(due to CSMA contention) and thus human detection rate will drop
as discussed earlier. }

\para{Non-human sources of motion.}  Smart homes and offices often
  have equipment that create motion even in the absence of human users.  One
  relevant question of interest is whether these machines will be detected by
  the attack as human movement, leading to false positives?  We experiment
  with a set of moving devices commonly seen in homes and offices, as well as
  pets, {\em e.g.} cats and dogs (see Table~\ref{table:machine}). 
  \shepherd{For example, robotic vacuums are placed 
  on the ground level and thus have minimal impact on the sniffed signals.}
  The only
  device to affect $\overline{\sigma_{aCSI}}$ in our tests is an oscillating
  fan. Yet its motion is highly periodic and consistent, making it easy to distinguish as
  non-human. We note that certain cats and dogs can also affect
  $\overline{\sigma_{aCSI}}$ with their movement, and their movement patterns
  can be hard to distinguish from human motion. Overall, our experiments show
  that the attack can eliminate all non-human sources of motion, except for pets.

\begin{table}[t]
  \centering
\resizebox{0.47\textwidth}{!}{
  \begin{tabular}{|c|c|c|}
    \hline 
Motion source   & \begin{tabular}[c]{@{}c@{}}Impact on \\
                    $\overline{\sigma_{aCSI}}$\end{tabular} & \begin{tabular}[c]{@{}c@{}}Distinguishable \\from human motion?\end{tabular} \\\hline 
Server internal cooling fan                     & No  &    -                                \\\hline
Standing fan                     & No &  - \\\hline

Oscillating fan                  & Yes  & Yes                                  \\\hline
Robot vacuum                     & No$^*$ &  No                            \\\hline
Cats or dogs & Yes   & No                                  \\        \hline                
\end{tabular}
}
\vspace{0.04in}
\caption{Impact of sources of non-human motion on our attack. (*) A robot
  vacuum only affects $\overline{\sigma_{aCSI}}$ 
  of an anchor in close proximity when the anchor is placed on the
  floor. } \label{table:machine}

\vspace{-0.2in}
\end{table}

\vspace{-0.1in}
\subsection{Evaluation of Bootstrapping}
\begin{figure}[t]
  \centering
\includegraphics[width=0.42\textwidth]{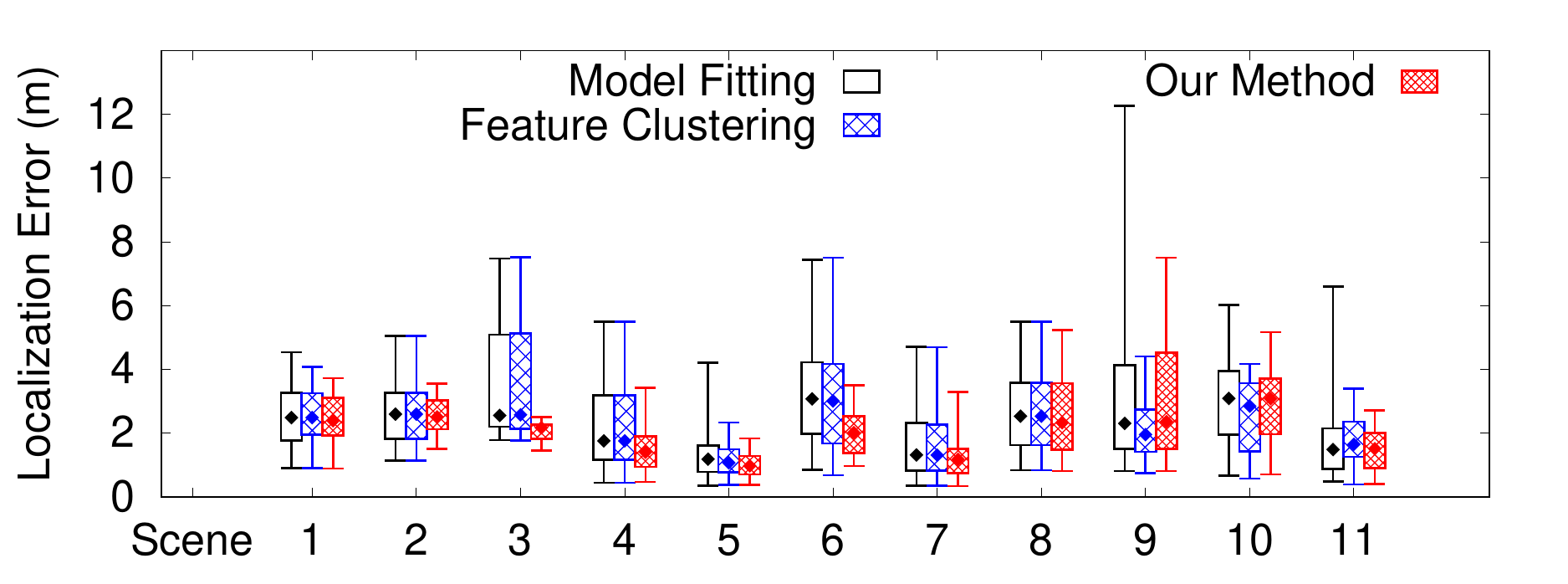}
\includegraphics[width=0.42\textwidth]{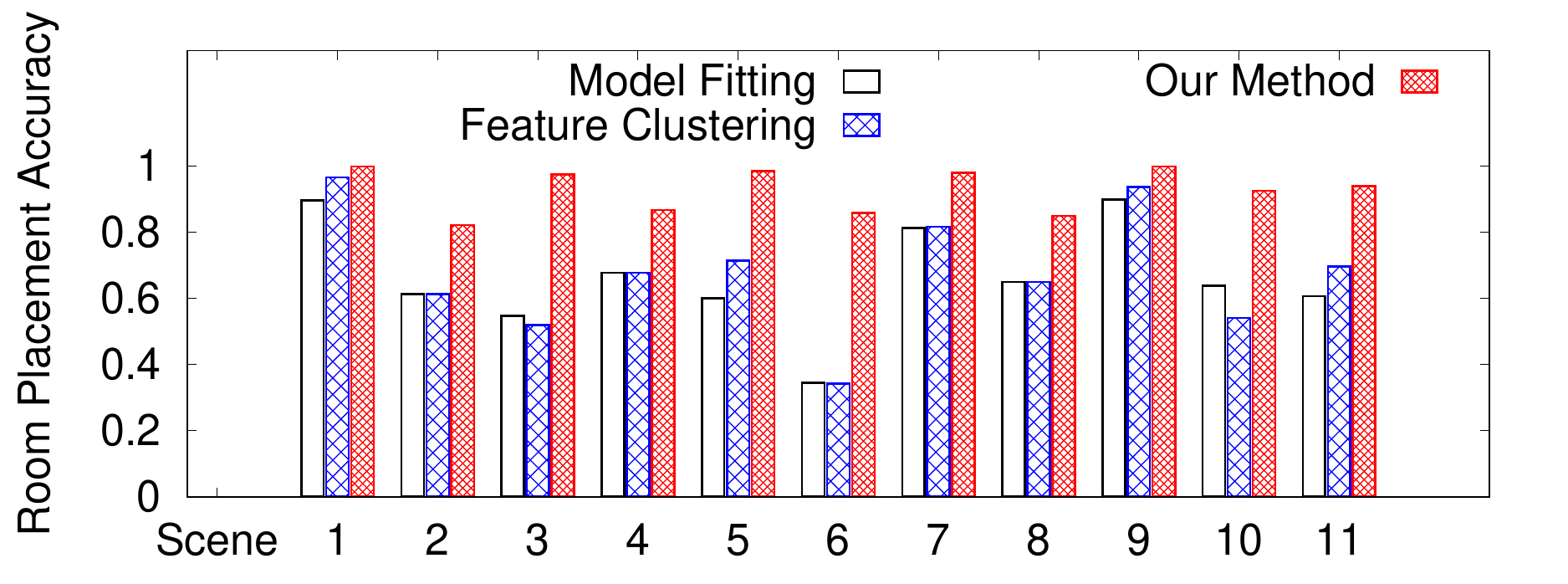}
\vspace{-0.1in}
    \caption{Bootstrapping performance: anchor localization accuracy in terms of
      absolute localization error (m) and room placement accuracy,
      per test scene.}
    \label{fig:roomestimation}
\end{figure}

We evaluate the bootstrapping phase (where the attacker detects and
locates anchors) via two metrics: {\em absolute localization
   error} which is the physical distance between the ground-truth location
 and the attacker-estimated location, and {\em room placement
   accuracy} which is the probability of placing an anchor to its
 exact room. 
Figure~\ref{fig:roomestimation} plots, for each test scene, the quantile distribution of the
absolute localization error and
the room placement accuracy.  We compare three systems: the model-fitting algorithm that uses all the measurements,
the feature-clustering based data filtering proposed
by~\cite{zhijing18hotmobile}, and our consistency-based data sifting
method.

\para{Gain of consistency-based data sifting.}  The results show that our proposed
data sifting method can largely improve anchor localization accuracy
compared to the two existing approaches.   Our method largely reduces
the localization error tail, and for more than 90\% of the
cases, the attacker places the anchor at the right room.  Our method
outperforms~\cite{zhijing18hotmobile} by using fine grained, scene-specific features to filter data. 

An interesting observation is that in scene 8--10,  our method
faces a similar (and even larger) absolute localization error than
feature clustering but produces
higher room placement accuracy. This is because our design directly 
analyzes the consistency of room placements, rather than raw localization errors.
Smaller raw localization error does not always translate into higher room
placement accuracy.

\para{Impact of anchor placements.} As expected, it is relatively
harder to accurately locate anchors placed at room boundaries, {\em
  e.g.}, those plugged into wall outlets.  In many cases, these boundary anchors create a dominant Monte Carlo cluster, but the data points in the
cluster map to either of the two neighboring rooms. Our current design
simply chooses the room with more data points, which
could lead to room placement errors. 

When the number of anchors is sufficiently large, the attacker can
minimize the impact of such boundary anchors in two ways. First, the
attacker can either use these boundary anchors 
``with caution'' or not use them at all.  Second, the attacker can
leverage past human sensing results to discovery any strong correlation
between anchors and adjust their room placements. We leave these to future work.

\para{Impact of anchor packet rate.}   The accuracy of
our proposed anchor
localization method is relatively insensitive to anchor packet
rate.   This is likely because  RSS (of static anchors) is relatively
stable over time. As long as the 
measurement trace covers $>$20m in distance and the RSS values
are between -75dB and -30dB without strong bias,  
we observe little difference in
localization (and room placement) accuracy.

\subsection{End-to-End Attack Evaluation}
Finally, we evaluate the
end-to-end performance of our attack, combining the bootstrapping and
continuous sensing phases. Like \S\ref{subsec:sensingeval}, we
consider 
the detection rate and false positive rate for the task of detecting and localizing human users to
their individual rooms.  
The results will include the impact of any misplaced anchors
during bootstrapping, or errors in detecting/localizing users during continuous
sensing.

\begin{table}[t]
\centering
\resizebox{1\columnwidth}{!}{
\begin{tabular}{cc|c|c|c|c|}
\cline{3-6}
                &           & \multicolumn{4}{c|}{\# of WiFi Devices Per Room} \\ \cline{3-6}
                                                &           & 1
                                                & 2          & 3
                                                & 4         \\ \hline
 \multicolumn{1}{|c|}{\multirow{2}{*}{Ours}} & DR    & 80.603\%    & 94.210\%    & 98.780\%    & 99.725\%   \\ \cline{2-6}
\multicolumn{1}{|c|}{}                          & FP & 3.595\%    &
                                                                    5.962\%
                                                            & 8.386\%
                                                & 10.719\%   \\ \hline
  \hline 
  \multicolumn{1}{|c|}{\multirow{2}{*}{LiFS}} & DR    & 14.153\%    & 26.381\%    & 36.954\%    & 46.033\%   \\ \cline{2-6}
\multicolumn{1}{|c|}{}                          & FP & 14.024\%
                                                & 14.077\%    &
                                                                14.493\%    & 15.064\%   \\ \hline
                                                
\end{tabular}
}
\vspace{0.04in}
\caption{End-to-end performance of our attack vs. LiFS, in terms of
  detection rate (DR) and false positive rate (FP).}
\label{table:e2e}
\vspace{-0.2in}
\end{table} 

Table~\ref{table:e2e} lists the detection rate and false positive rate  for our attack design and those achieved via LiFS~\cite{lifs16mobicom}. We
also vary the number of WiFi anchor devices in each room.
Compared to the results in Table~\ref{table:userdetection} assuming accurate anchor room
placement, the end-to-end attack sees minor drop in both recall
and precision, especially with more anchors per room. Despite using a
passive, minimally equipped attacker device,  our attack still
achieves high human sensing accuracy, {\em e.g.\/}, 99.7\% detection
rate at 10.71\% false positive rate. 

The impact of anchor localization errors is much more
visible on LiFS, whose detection rate drops to 36.954\% and 46.033\% even with
3 and 4 anchors in each room, respectively.  Overall, we see that
while both using the same aCSI values per
anchor,  our proposed passive human sensing largely outperforms
LiFS by not requiring precise anchor location to model signals on the direct propagation path.

\section{Defenses}
\label{sec:defense}
We now explore robust
defenses against our proposed attack and other passive sensing
attacks.  Our design insight is that attack
effectiveness depends heavily on the quantity and quality
of the 
WiFi signals captured by the sniffer.  Thus a defense reducing the amount
of WiFi signal leakage to external sniffers or adding inconsistency to WiFi
signals could render the attack ineffective.

\subsection{MAC Randomization} 
The first solution coming 
to mind would be 
{\em MAC address randomization}, a well-known method for protecting mobile devices
against tracking.  Since the attack sniffer uses MAC address to isolate
signals of anchors, MAC randomization can disrupt both
bootstrapping and continuous sensing phases.  However, recent 
work has shown that MAC randomization is disabled on most
devices ($<$3\% of adoption rate so far)~\cite{randommacspread} and can be
easily broken to reveal the real MAC
address~\cite{macrandomization,randommac100perc}.  We note that Android
9.0 Pie switches to per-network MAC
randomization~\cite{randommacAndroidP},  which does not apply any MAC
randomization to static WiFi devices. Thus MAC randomization is not a
plausible defense against our attack.

\subsection{Geofencing WiFi Signals}
Geofencing bounds signal propagation to  reduce or eliminate WiFi signals accessible to the
adversary. In our attack, when the area with
a signal in our walking trace reduces from 25--50 meters to  10 meters or less,  the anchor localization error increases
significantly:  raw
errors more than double, and anchor room placement accuracy drops from 92.6\% to
41.15\%.

Geofencing is also extremely difficult to deploy and configure.  The simplest form
is to reduce the anchor's transmit power, which is almost always
undesirable since it degrades connectivity. Another option is to 
equip WiFi devices with directional antennas, limiting signal spatial
coverage.  This is also undesirable as it requires upgrading to equipment with higher cost and larger
form factors, as well as carefully configuring antenna directionality.
Finally, the extreme solution is to block RF  propagating beyond
property walls by painting these walls with electromagnetic shielding paints.
This is again impractical, since it blocks incoming WiFi/cellular 
signals.

If the area accessible to the attacker is limited, a potential solution is to customize WiFi signal
coverage using 3D fabricated
reflectors~\cite{xiongxi17buildsys} or 
backscatter arrays~\cite{kyle19nsdi} that create noise towards the
area.  Yet both remain open research problems.

\subsection{WiFi Rate Limiting} 
While geofencing reduces spatial leakage of WiFi signals,  rate limiting
reduces their temporal volume. When anchors transmit less signals over time,
the sniffer will not have sufficient data to compute
$\overline{\sigma_{aCSI}}$. 
Results in \S\ref{sec:eval}  show that reducing anchor packet rates
does lower the detection rate (when using a single anchor) but can be
compensated by aggregating the results of multiple anchors. 

In practice, rate limiting is undesirable for
most 
network applications. As shown in
Table~\ref{tbl:devices_summary}, many WiFi devices, when idle, already transmit at more than 2pps. It is hard
to rate limit further, rendering the defense ineffective. 

\subsection{Signal Obfuscation: Existing Designs}
{\em Signal obfuscation}  adds noise to
WiFi signals, so the adversary
cannot accurately localize anchors or detect user motion.  Existing
works have proposed both temporal and spatial obfuscations against
RF sensing~\cite{zhijing18hotmobile, phycloak16nsdi}.

 In {\em
  temporal obfuscation}, WiFi devices (anchors) change transmit
power randomly over time,  injecting artificial  noises to
signals seen by the sniffer.  Doing so, however, requires upgrading 
commodity WiFi devices to equipment with much higher cost
and energy consumption. Also a more resourceful adversary
can counter by deploying an extra static sniffer (during bootstrapping) to infer the injected signal power changes and 
remove them from the signal traces, as shown by~\cite{zhijing18hotmobile}.

In {\em spatial obfuscation}, a recent work~\cite{phycloak16nsdi} shows that by deploying a
full-duplex radio near each 
anchor $x$, one can obfuscate $x$'s signal phase, RSS, CSI, and Doppler
shift seen by any nearby sniffers with a single antenna.  But full-duplex 
radios are of high cost, and there is no
commodity product on the market.  On the other hand, defending against
our attack only
needs to obfuscate RSS and aCSI, collected by the sniffer. 

\subsection{Proposed: AP-based Signal Obfuscation}
The above four immediate defenses are either ineffective or
impractical.  Instead, we propose 
a practical defense where
the WiFi access point (AP) actively injects customized cover traffic for any of its
associated WiFi device $w$ that is actively transmitting.  We design
this defense to produce large ambiguity to the attack in two
steps. {\em  First}, our defense adds noise to the attacker's RSS measurements,
so that during
bootstrapping, the attacker will place most of the anchors to the
wrong room and even outside of the property.  {\em Second}, our
defense largely reduces (and even removes) the 
$\overline{\sigma_{aCSI}}$ gap between no human presence and human
motion,  such that the attacker is unable to identify human motion.

\para{AP inserting customized cover signal.} As soon as the AP
detects a transmission from $w$, it estimates $w$'s transmission rate
$T_w$ and injects a cover traffic stream with the rate of  $\rho 
T_w$, at a randomized
power level and with $w$'s MAC address. 
$\rho$ is the injection rate,
a system parameter. Since the AP limits its cover traffic stream
to be proportional to $w$'s throughput, the CSMA protocol will randomly interleave packets
from the two streams together. In the worst case ($\rho T_w$ is at or higher than
available channel throughput), the cover traffic will reduce $w$'s effective
throughput by $1+\rho$.

The insertion of ``fake'' packets requires a careful design, so that it
disrupts the attack rather than creating obvious ``anomalies'' or heavily
affecting the WiFi network. In particular, the AP configures the sequence numbers of fake
packets to (partially) interleaved with those of real packets, so that the
attacker is unable to separate the two streams based on sequence number and
packet arrival time. 
\shepherd{When sending fake packets, the AP's transmit power is
  randomized but close to that of $w$, so the mixed traffic follows
  natural (and complex) multipath signal variation. This makes it hard
  to distinguish real and fake packets from signal strength values alone.}

Finally, this defense can be deployed on today's WiFi
APs that support transmit power adaptation with minor changes.  The
major overhead is the extra consumption (a factor of $\rho$) of bandwidth and energy at the AP.

\para{Results: Impact on bootstrapping.} With this defense,  the attacker's RSS measurements of anchor $w$ will display
fluctuations, tricking the sniffer to think that $w$ is
moving and not use it as an anchor.  Even if the adversary
assumes $w$ is stationary, the noisy RSS measurements (even after our
data sifting) will lead to
inaccurate room placement. 

When evaluating this defense, we consider both our original attacker
(with one smartphone) 
and an {\em advanced} attacker who adds an extra stationary sniffer and applies
RSS signal
subtraction to detect and remove any ``injected'' signal
variations~\cite{zhijing18hotmobile}.   We configure our defense where the AP injects cover traffic of
$\rho$ with power randomization in the $10dB$ range. 
For both
attackers,  $\rho$=5\% is sufficient to drop the accuracy of anchor
room placement from 92.6\% (without our defense) to 35.71\%, except for the anchors close
to the AP (in the same room).  As we further increase $\rho$, the
attacker will map most of the detected anchors to the AP's room.

\begin{figure}[t]
\centering
  \includegraphics[width=0.23\textwidth]{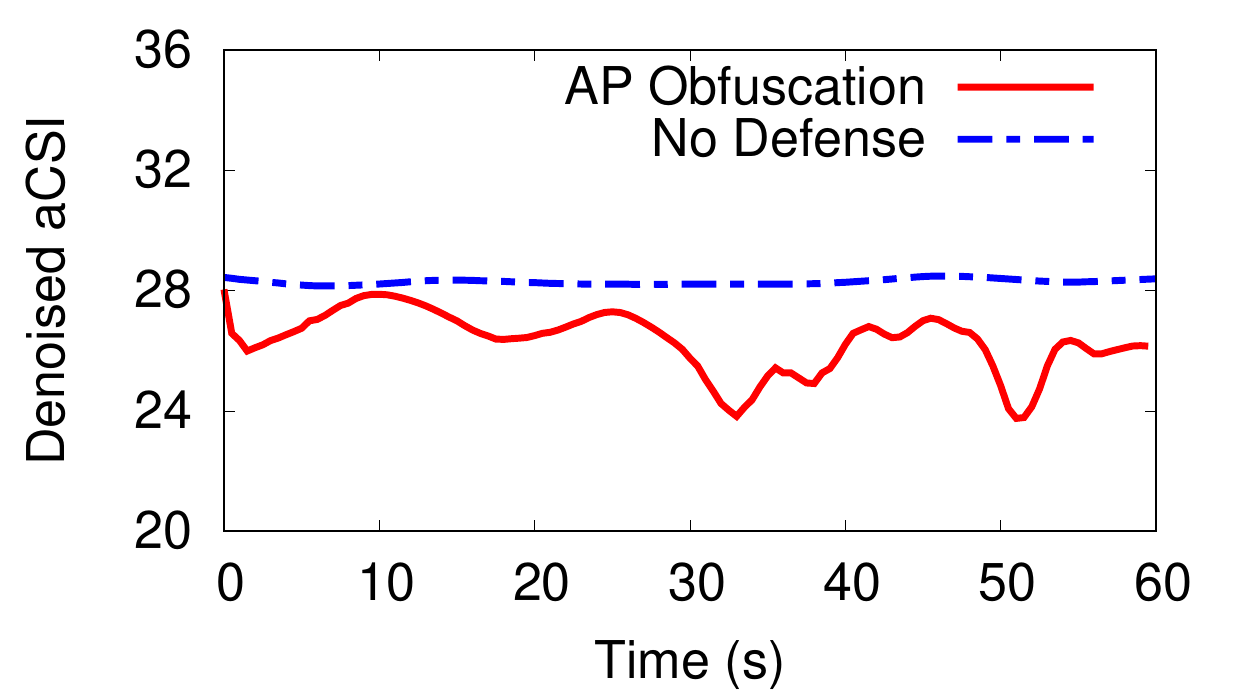}
  \includegraphics[width=0.23\textwidth]{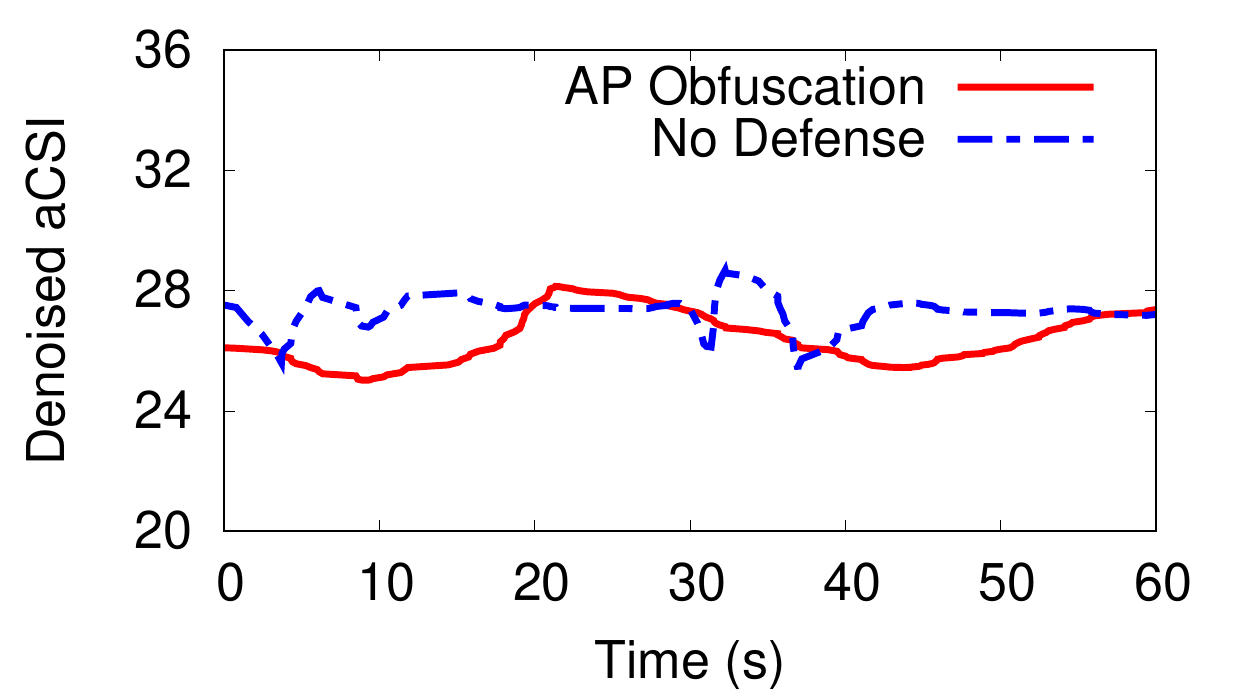}
\subfigure[User not present]{
  \includegraphics[width=0.225\textwidth]{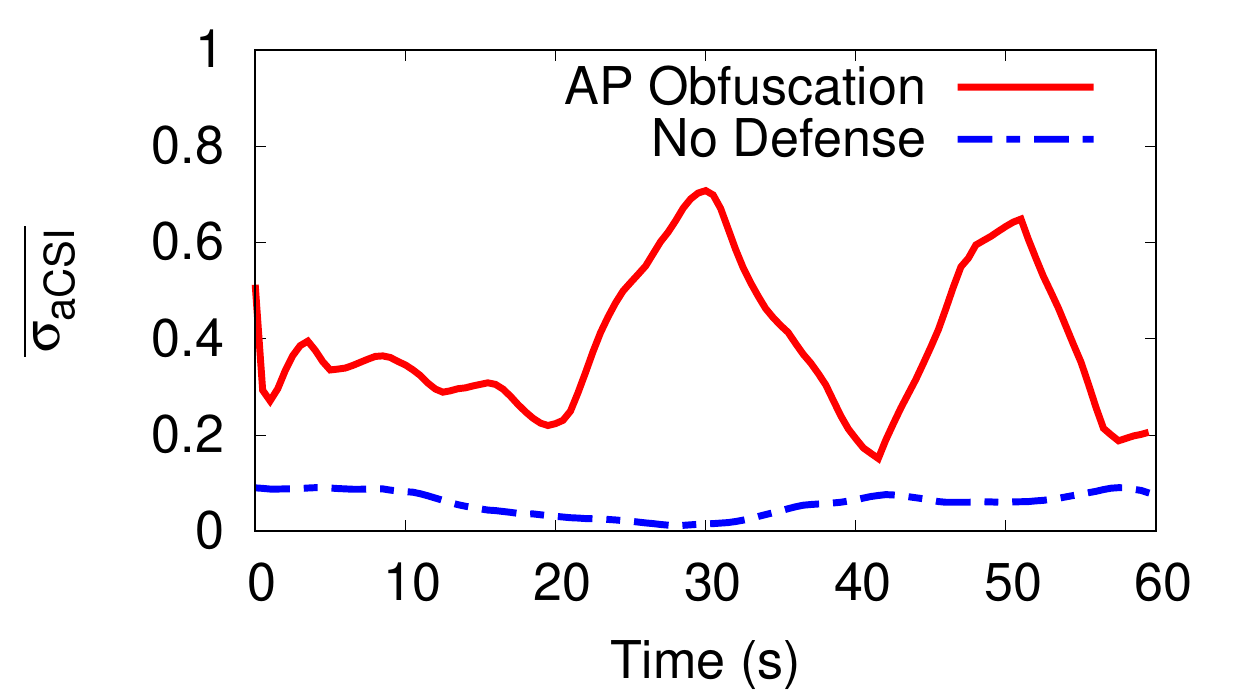}
   \label{fig:defense_noone}
     }
\subfigure[User in motion]{
  \includegraphics[width=0.225\textwidth]{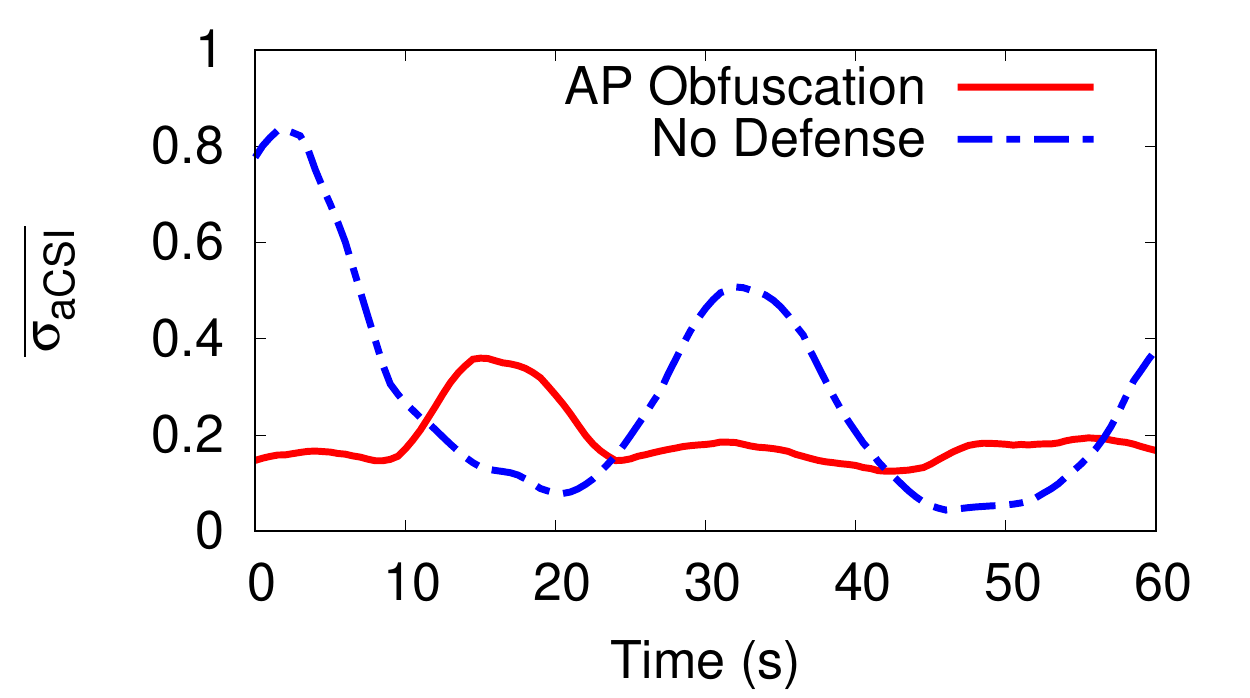}
  \label{fig:defense_moving}
}
\caption{aCSI and $\overline{\sigma_{aCSI}}$ with and without AP
  based signal obfuscation. }
 \end{figure}

\para{Results: Impact on continuous sensing.} As the attacker sniffer
calculates $\overline{\sigma_{aCSI}}(w)$ on randomly interlaced packets
sent by anchor $w$ and the AP,  the value of 
$\overline{\sigma_{aCSI}}(w)$ with no human presence will
increase significantly. Figure~\ref{fig:defense_noone} shows a sample 
trace of aCSI (of a sub-carrier) and $\overline{\sigma_{aCSI}}$ with
and without the defense.  We see that our defense can effectively
elevate the threshold $\sigma_{p}(w)$ for human presence detection.
More importantly,  the defense has much less impact on 
$\overline{\sigma_{aCSI}}(w)$ when there is actual human movement near the
anchor $w$.  The sample traces in Figure~\ref{fig:defense_moving} show
that $\overline{\sigma_{aCSI}}(w)$ actually drops (below the threshold) when using the
proposed defense.  In this case, human movement will not
trigger any anchor in proximity, for both the original and the
advanced attackers (who deploy an extra sniffer).

Table~\ref{table:sensing_impact} lists the attack performance with our
proposed defense ($\rho$=30\%) and without any defense. We first
consider the case where the attacker manages to obtain perfect anchor
room placement.  In this case, our defense increases the false positive rate from 7.9\% to
48.28\% while dropping the detection rate to 78.776\%.  Next, we
consider the end-to-end attack scenario where the attacker performs
both bootstrapping and continuous sensing.  Our defense drops the detect rate down to 47.48\% while increasing the 
false positive rate to  49.5\%. These results apply to both the original
attacker and the advanced attacker.  Such ambiguity renders the attack useless in practice.

\begin{table}[h]
\resizebox{\columnwidth}{!}{
\begin{tabular}{l|c|c|c|c|}
\cline{2-5}
\multirow{2}{*}{}
  & \multicolumn{2}{c|}{False positive rate} & \multicolumn{2}{c|}{Detection rate} \\ \cline{2-5} 
                                                                                         &  No
                                                                                           defense
                                     & AP obf    & No defense   & AP obf \\ \hline
\multicolumn{1}{|l|}{\begin{tabular}[c]{@{}l@{}} knowing anchor 
                       \\ room placement\end{tabular}} & 7.935\%      & 48.284\%       & 99.988\%     & 78.776\%     \\ \hline
\multicolumn{1}{|l|}{\begin{tabular}[c]{@{}l@{}}end-to-end \\ attack\end{tabular}}                                                   & 10.719\%      & 49.598\%       & 99.725\%     & 47.481\%     \\ \hline
\end{tabular}
}
\vspace{0.04in}
\caption{The attack performance under AP-based signal
  obfuscation (best performance out of the original and the advanced attack with an extra sniffer). }
\label{table:sensing_impact}
\end{table}

\shepherd{
\para{Possible countermeasures.} 
To overcome our proposed defense,
the attacker must find ways to distinguish the obfuscation 
packets sent by AP from the original packets sent by an anchor $w$. As
discussed earlier, doing so using packet sequence number and arrival
time is infeasible due to our packet injection method.   Doing so at the
network traffic level is also difficult, since packet contents are encrypted, and we can
shape traffic to resist traffic
identification by attackers~\cite{covertraffic}.   Finally, it is also difficult to
separate the two streams using physical layer characteristics, because
doing so requires much more sophisticated and bulky hardware. 
One option is to analyze per-symbol aCSI/RSS patterns. This is infeasible using commodity
WiFi chips, as they only report per-packet aCSI/RSS values. Another
option is to use a large antenna array (MIMO with at least 4--6 antenna
elements, each separated by 6.25cm) to distinguish signals sent by $w$
from those sent by the AP, since they come from different directions.  The
resulting sniffer ($>$31cm in length) would be conspicuous and easily raise suspicion.  
}

\section{Related Work}
\label{sec:related}
\spara{Human sensing by snooping signals.} We categorize 
existing works into five groups. 
\shepherd{
The first group applies {\em traffic analysis} to infer user presence and status in a home/office from
their network traffic~\cite{videosniffing16infocom,privacy14iccst,fanzhang11wisec,dewicam18asiaccs, gtid15fingerprint, homesnitch19wisec,peekaboo18}.
It requires strong knowledge on device
behaviors and can be easily countered by 
sending cover traffic, applying encryptions and traffic shaping.
In contrast, our attack remains effective even when all network-level defenses are deployed, as long as WiFi devices still transmit packets.
}

The second group uses ``specialized signals'' such as RFID~\cite{rfid17survey}, 
visible light~\cite{litell16mobicom,lightsense16mobisys},
and acoustic~\cite{cat16mobicom, shyam17ubicomp}, that often correlate with 
human motion.  But existing solutions require control of  
transmitters inside or outside of the target property, which is 
infeasible under our attack model. 

The third group builds {\em fingerprints} of each predefined target location
and/or activity, based on either aCSI~\cite{NandakumarKG14,jianliu14mobicom,passivecsi18NaNA}, CSI~\cite{wisee13mobicom, weiwang15mobicom}, RSS~\cite{widet18mswim,srinivasan08ubicomp,nuzzer13youssef,rfsensing14tomc}, or
raw signals~\cite{motionfi18infocomm}.  Since the attacker under our
model has no knowledge of the target users and access to the target
property,  building fingerprints becomes infeasible. 

The fourth group uses advanced radio hardware (laptops or USRPs with
antenna arrays or directional antennas) that communicate with the
anchors inside the target property.  This allows the sniffer to 
measure fine-grained CSI values
(both amplitude and phase)\shepherd{~\cite{freesense16globecom}}, and use
them to calculate AoA and doppler frequency shift (DFS) to detect human
motion~\cite{wideo15nsdi,widar2_18mobisys,weiwang15mobicom,
  yousefi2017survey, activitytrain18mobicom, passiveradar12}.  Our
attack differs by using a passive sniffer with a single
antenna, which does not
communicate/synchronize with the anchors.  In this case, the sniffer
cannot infer CSI phase, AoA or DFS.

The final group detects user motion using passive
sniffers
\shepherd{to collect and analyze physical RF signals}~\cite{banerjee14wisec,passiveradar12, lifs16mobicom}. As
discussed earlier, both~\cite{banerjee14wisec, lifs16mobicom} target 
user motion that disturbs the direct propagation path, 
requiring precise locations of the anchors. ~\cite{passiveradar12} uses multiple sniffers with bulky
directional antennas to compute doppler shift of user motion.  The sensing method used by our attack falls into this
category, but targets multipath signal
propagation from each anchor to the sniffer. We design a new
aCSI variance model to reliably detect user motion,  eliminating the need
for precise anchor location and antenna array at the sniffer.

\spara{Passive transmitter localization.}  Existing works often
leverage bulky receivers with
multiple antennas~\cite{adib2013, spotfi15sigcomm, multipathtriang18mobisys,
  mostofi18ipsn,arraytrack13nsdi,vrwifitracking17cvpr} to estimate signal AoA, and applies triangulation across receivers to
derive target location.
Our anchor localization (during bootstrapping) uses a compact smartphone with a single
antenna, and applies passive localization 
that fits spatial RSS measurements to a propagation
model~\cite{ariadne06mobisys,zhijing18hotmobile,wigem11conext}.   Our key contribution is the data sifting algorithm that
identifies good RSS samples as input to the model fitting. 

\spara{Defense against RF sensing.} Existing
works~\cite{carving12asiaccs, phycloak16nsdi, ijam10, waveforming15ccs}
defend against eavesdropping on a transmitter by a jammer 
transmitting simultaneously, preventing the attacker from decoding packets
or estimating CSI/AoA.  
This requires precise synchronization between the
transmitter and the
jammer~\cite{defense16winet}  or a
high-cost full-duplex obfuscator~\cite{phycloak16nsdi}. Our defense
uses AP to insert fake packets (rather than transmitting simultaneously), 
which is easy to deploy and effective against
our attack.

\section{Conclusion}
\label{sec:discussion}
\readmore{

  Our work shows that the ubiquity of WiFi devices has
  an unexpected cost: reflected or blocked RF transmissions leak
  information about our location and activities. We describe a set of
  low-cost, stealthy reconnaissance attacks that can continuously monitor and
  locate human motion inside a private property, turning WiFi
  devices inside into motion sensors.  All this is done without
  compromising the WiFi network, data packets or devices, and only requires a
  commodity WiFi sniffer outside of the property.  We validate the attack on
  a variety of real-world locations, and develop a new effective defense
  based on carefully tuned WiFi signal obfuscation by APs.

  We believe our work points to the potential of more powerful information
  leakage attacks via passive RF reflections. With more sophisticated signal
  processing techniques (and potentially new hardware), much more might be
  learned from the way ambient RF signals interact with our bodies and
  surroundings. We are pursuing this line of research to both better
  understand these attacks and to develop defenses to better safeguard
  our security and privacy. 

}

\section*{Acknowledgment}
We thank our shepherd Earlence Fernandes and
the anonymous reviewers for their feedback.  We also thank
Vyas Sekar and Fadel Adib for their feedback on the early version of
this work. 
This work is supported in part by the National Science Foundation
grants CNS-1923778 and CNS-1705042.
Any opinions, findings, and conclusions or recommendations
expressed in this material do not necessarily reflect the views of any
funding agencies.

\bibliographystyle{IEEEtranS} 
\bibliography{advloc}

\end{document}